\documentclass[useAMS,usenatbib,usedcolumn]{mn2e}
\usepackage{epsfig}
\usepackage{amsmath}
\usepackage{amssymb}
\usepackage{mathrsfs}

\title[Gravitational Lensing in CANDELS \& the XDF]{The Impact of Strong Gravitational Lensing on Observed Lyman-Break Galaxy Numbers at $4 \leq z \leq 8$ in the GOODS and the XDF Blank Fields}

\author[Barone-Nugent et al.]{R. L. Barone-Nugent$^{1}$, J.S.B Wyithe$^{1}$, M. Trenti$^{1,2}$, T. Treu$^{3}$, P. Oesch$^{4}$, \newauthor R. Bouwens$^{5}$, G. D. Illingworth$^{6}$, K. B. Schmidt$^{3}$\\
$^{1}$ School of Physics, University of Melbourne, Parkville, Victoria, Australia\\
$^{2}$ Kavli Institute for Cosmology and Institute of Astronomy, University of Cambridge, Cambridge, UK\\
$^{3}$ Department of Physics, University of California, Santa Barbara, CA 93106-9530, USA\\
$^{4}$ Department of Astronomy, Yale University, New Haven, CT 06511, USA\\
$^{5}$ Leiden Observatory, Leiden University, NL-2300 RA Leiden, Netherlands\\
$^{6}$ UCO/Lick Observatory, University of California, Santa Cruz, CA 95064, USA\\
Email: robertbn@student.unimelb.edu.au}

\begin{document}
\date{\today}

\maketitle

\label{firstpage}

\begin{abstract}

Detection of Lyman-Break Galaxies (LBGs) at high-redshift can be affected by gravitational lensing induced by foreground deflectors not only in galaxy clusters, but also in blank fields. We quantify the impact of strong magnification in the samples of $B$, $V$, $i$, $z$ $\&$ $Y$ LBGs ($4\lesssim z \lesssim8$) observed in the XDF and GOODS/CANDELS fields, by investigating the proximity of dropouts to foreground objects. We find that $\sim6\%$ of bright LBGs ($m_{H_{160}}<26$) at $z\sim7$ have been strongly lensed ($\mu>2$) by foreground objects. This fraction decreases from $\sim 3.5\%$ at $z\sim6$ to $\sim1.5\%$ at $z\sim4$. Since the observed fraction of strongly lensed galaxies is a function of the shape of the luminosity function (LF), it can be used to derive Schechter parameters, $\alpha$ and $M_{\star}$, independently from galaxy number counts. Our magnification bias analysis yields Schechter-function parameters in close agreement with those determined from galaxy counts albeit with larger uncertainties. Extrapolation of our analysis to $z\gtrsim 8$ suggests that future surveys with JSWT, WFIRST and EUCLID should find excess LBGs at the bright-end, even if there is an intrinsic exponential cutoff of number counts. Finally, we highlight how the magnification bias measurement near the detection limit can be used as probe of the population of galaxies too faint to be detected. Preliminary results using this novel idea suggest that the magnification bias at $M_{UV}\sim -18$ is not as strong as expected if $\alpha\lesssim -1.7$ extends well below the current detection limits in the XDF. At face value this implies a flattening of the LF at $M_{UV}\gtrsim-16.5$. However, selection effects and completeness estimates are difficult to quantify precisely. Thus, we do not rule out a steep LF extending to $M_{UV}\gtrsim -15$.

\end{abstract}

\section{Introduction}
\label{section:introduction}
\indent Surveys of Lyman-Break Galaxies (LBGs) during the epoch of reionization aim to make a census of high-redshift galaxies and to estimate the available ionizing photon budget for Reionization \citep{bouwens2014uv, finkelstein2014evolution, schmidt2014luminosity, schenker2013uv, robertson2013new, mclure2013new, finkelstein2012candels, bradley2012brightest, oesch2012bright, bouwens2011ultraviolet, castellano2010bright, bouwens2008z, khochfar2007evolving}. These surveys, however, may provide an increasingly skewed view of the early Universe as redshift increases. The observations are complicated by gravitational lensing \citep{wyithe2011distortion}, which affects the observed luminosities and surface density of high-redshift LBGs. Along random lines of sight, the probability of significant magnification and multiple images from gravitational lensing is $\sim0.5\%$ \citep{barkana2000high, comerford2002constraining, wyithe2011distortion}. Furthermore, high-redshift luminosity functions have been shown to have very steep faint end slopes \citep[$\alpha\sim-1.6$ at $z\sim4$ to $\alpha\sim-2.0$ at $z\sim8$, although the uncertainties are large, e.g.][]{bouwens2014uv, schmidt2014luminosity}, which results in a bias leading to an enhanced probability of gravitational lensing over random lines of sight. This so-called \textit{magnification bias} is further enhanced for flux limits at magnitudes brighter than $M_{\star}$ where number counts drop exponentially. Consequently, bright LBGs become much more likely to have been gravitationally lensed than random lines of sight \citep{wyithe2011distortion}. The strongly-lensed fraction of LBGs at $z\sim7-8$ brighter than $M_{\star}$ is expected to be $\sim10\%$.\\
\indent The amplitude of magnification bias is a function of $M_{\star}-M_{\textrm{lim}}$ and $\alpha$ \citep{pei1995magnification, wyithe2011distortion}, where $M_{\mathrm{lim}}$ is the survey flux limit. Therefore, quantifying the amount of magnification bias offers a direct probe of the luminosity function down to, and below, current survey detection limits \citep{mashian2013constraining}. Alternative approaches to quantifying the LF beyond detection limits exist, such as targeting massive galaxy clusters as gravitational lenses \citep{atek2014new, alavi2014ultra, yue2014ultra, ishigaki2014hubble} and assessing the noise characteristics of the background \citep{calvi2013constraining}.\\
\indent Magnification bias is expected to significantly skew the observed bright-end of the luminosity function at very high redshifts (see Fig. 3 in \citealt{wyithe2011distortion}). The effect is intriguing in light of the $z\sim7$ LF, which has been observed to both agree with the exponential cutoff in the Schechter parametrization \citep{bouwens2014uv}, and also to decline less steeply than a Schechter function \citep{bowler2014bright}. Previous studies of high-redshift galaxies have argued that gravitational lensing has not significantly affected their luminosity functions \citep{mclure2006discovery}, while others have made slight corrections in the observed luminosities of LBGs due to gravitational lensing \citep{bowler2014bright}. However, even though the sample sizes analyzed in recent studies are large, the numbers of bright galaxies, where the magnification bias effects will be most apparent, remain small.  Thus observational verification of any changes in slope at the bright end remains to be determined.\\
\indent Identifying and confirming individual cases of strong gravitational lensing of LBGs at $z\gtrsim4$ is made difficult due to sources appearing faint and small. Elongation in the observed LBG due to lensing is difficult to detect due to their small observed size compared with the resolution of the telescope. Secondary images are very difficult to observe, as they will be less magnified than the primary image, and hence be extremely faint \citep{barone2013candidate}. Secondary images will also appear closer to the deflector than the primary image, making it more difficult to disentangle them from the deflector galaxy light than the primary image. \citet{wyithe2011distortion} calculated the probability of detecting a secondary image to be $\approx10\%$, where the bright image of a galaxy is one magnitude above the survey limit.\\ 
\indent In this paper, we adopt a statistical approach to detect gravitational lensing of high-redshift galaxies using the largest samples of LBGs at $4\leq z \leq 8$ \citep{bouwens2014uv}. We assess the likelihood of lensing for each individual LBG in homogenous samples at $z\sim4$, $z\sim5$, $z\sim6$ and $z\sim7-8$ in order to infer the total expected lensed fraction at a range of flux limits. In Section \ref{section:data} we describe the data used in our analysis. In Section \ref{section:FJR} we derive the Faber-Jackson relation \citep{faber1976velocity} of foreground galaxies that we will use in our analysis. Section \ref{section:method} describes our method of prescribing a likelihood of lensing to each LBG. Section \ref{section:results} describes our lensing results, and in Section \ref{section:magbias} we assess the magnification bias and the consequences for the LF beyond current survey limits. In Section \ref{mag-lumfunc} we assess the observational effects of magnification bias on the LF and in Section \ref{Jdrops} we present the strong lensing likelihoods of existing $z\sim9-10$ LBGs. In Section \ref{section:conclusion} we conclude. We refer to the HST F435W, F606W, F775W, F850LP, F105W, F125W, and F160W bands as $B_{435}$, $V_{606}$, $i_{775}$, $z_{850}$, $Y_{105}$,  $J_{125}$ and $H_{160}$. Throughout this paper we use $\Omega_{M}=0.3$, $\Omega_{\Lambda}=0.7$ and $H_{0}=70$ km/s/Mpc, and all magnitudes are in the AB system \citep{oke1983secondary}.

\section{Data}
\label{section:data}
\indent The analysis presented in this paper makes use of the wide-area, ultra-deep observations of the XDF/UDF and GOODS (from the XDF, ERS and CANDELS programs) \citep{illingworth2013hst, windhorst2011hubble, grogin2011candels, koekemoer2011candels}. The observations cover the $4.7$ arcmin$^{2}$ area of the XDF, which reaches $\sim30$ mag at $5\sigma$, the $126$ arcmin$^{2}$ of the GOODS Deep fields, which reach $\sim28.5$ mag at $5\sigma$, and the $115$ arcmin$^{2}$ of the GOODS Wide fields, which reach $\sim27.7$ mag at $5\sigma$. The catalogues were constructed to identify Lyman-Break galaxies from $z\sim4$ to $z\sim8$ using a colour-colour criteria (see details in Section 3.2.2 of \citealt{bouwens2014uv}). LBGs `drop out' in the $B_{435}$, $V_{606}$, $i_{775}$, $z_{850}$ and $Y_{105}$ for LBGs at $z\sim4$, $z\sim5$, $z\sim6$, $z\sim7$ and $z\sim8$, respectively. The catalogues include $5867$, $2108$, $691$, $455$ and $155$ LBG candidates at $z\sim4$, $z\sim5$, $z\sim6$, $z\sim7$ and $z\sim8$ respectively.\\
\indent We use the 3D-HST photometric catalogue of the CANDELS area \citep{skelton20143d, brammer20123d} in order to model foreground objects as potential gravitational lenses. The 3D-HST survey covers all of the CANDELS fields, with spectroscopy compiled from the literature. We utilize the spectroscopic redshifts of foreground objects when available, and otherwise rely on photometric redshifts obtained using $\tt{EAZY}$ \citep{brammer2008eazy} which are based on deep multiband observations \citep{erben2008cars, taniguchi2007cosmic, grogin2011candels, koekemoer2011candels, whitaker2011newfirm, mccracken2012ultravista, bielby2011wircam, brammer20123d, ashby2013seds, barmby2008catalog}.\\
\indent We calibrate a redshift-dependent Faber-Jackson relation \citep{faber1976velocity} in Section \ref{section:FJR} using early-type galaxy data from \citet{treu2005assembly}, \citet{auger2009sloan}, and \citet{newman2010keck}. The galaxies in these samples have published spectroscopic redshifts, velocity dispersions and rest-frame $B$-band magnitudes.\\

\section{The Faber-Jackson Relation}
\label{section:FJR}
\begin{figure*}
\includegraphics[scale=0.212]{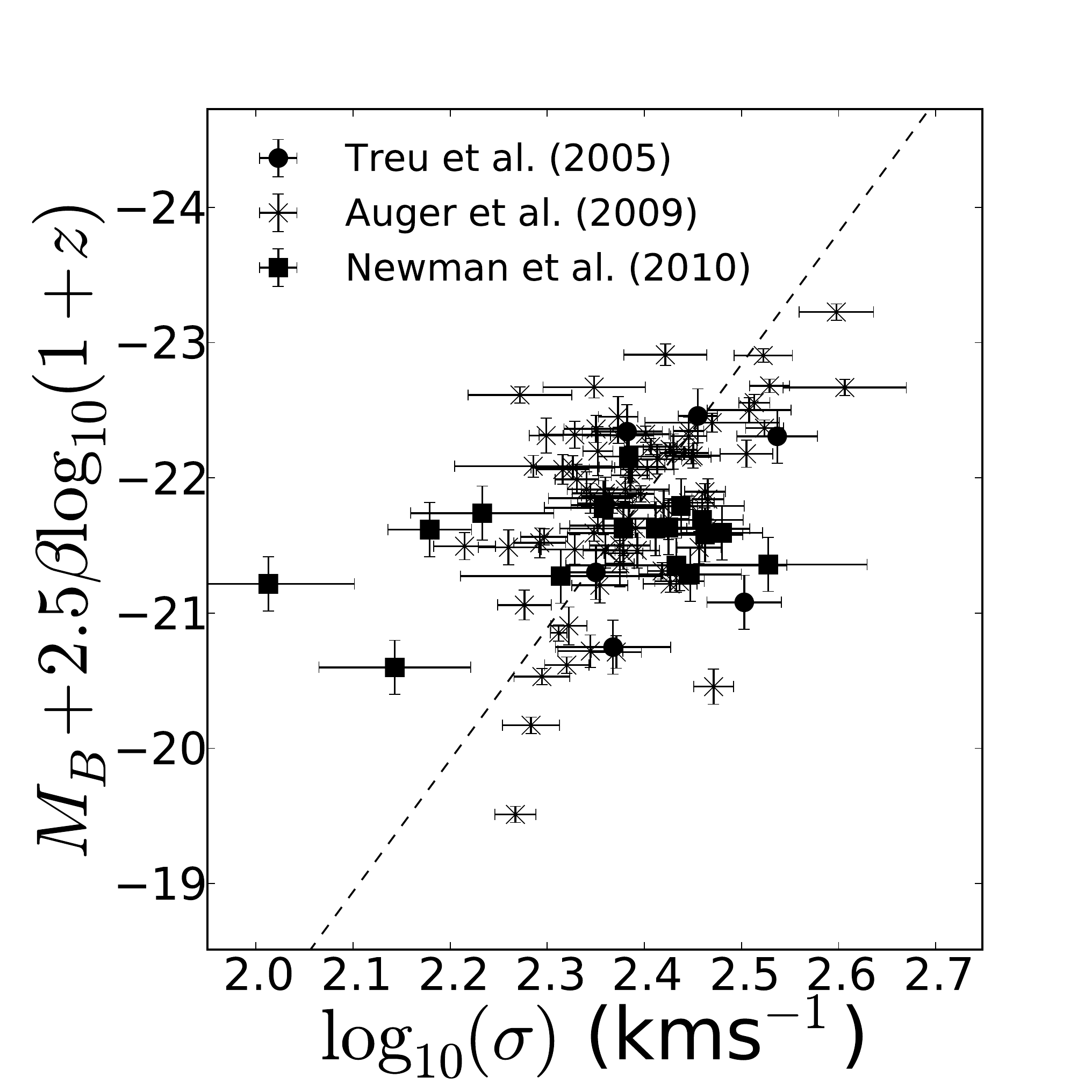}
\includegraphics[scale=0.212]{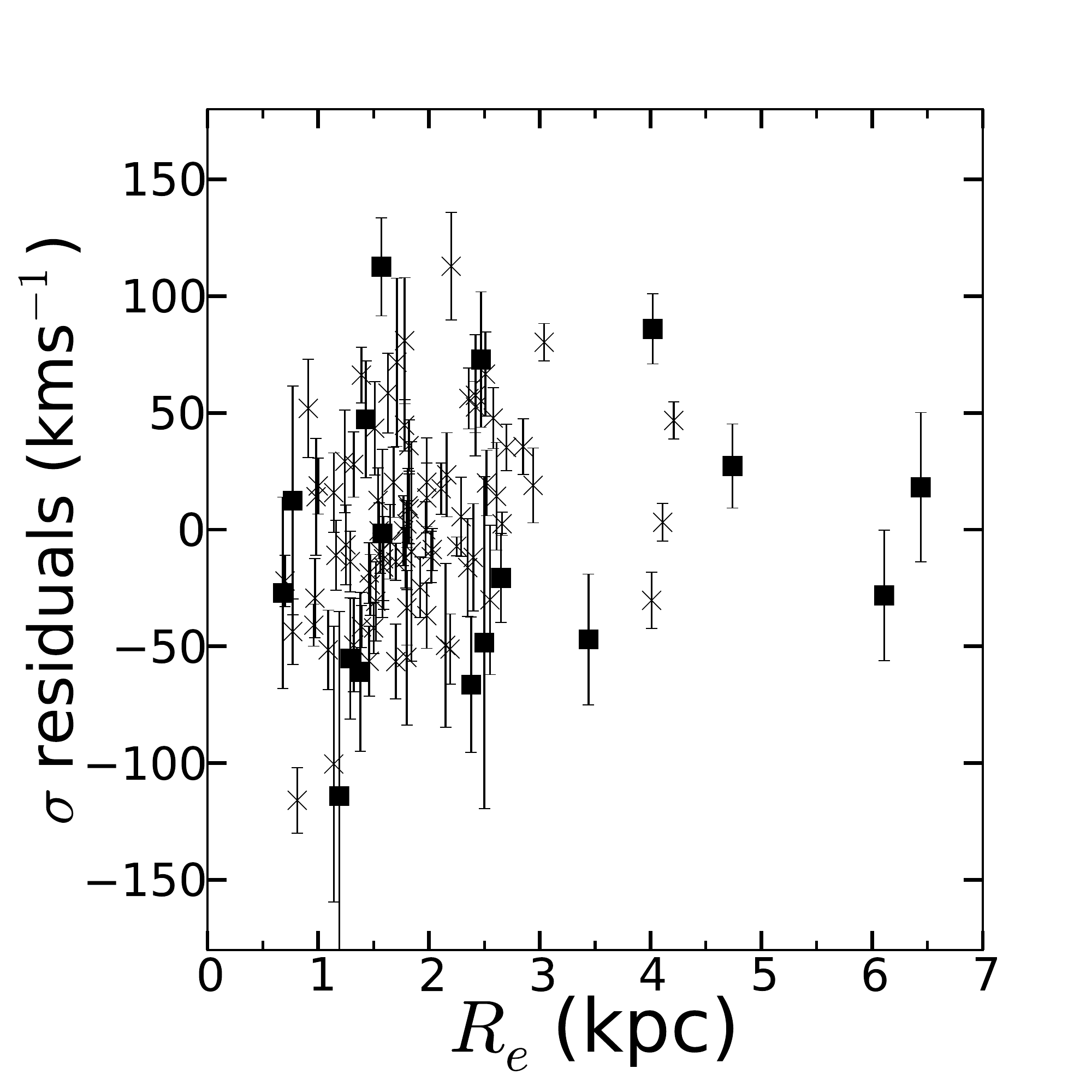}
\includegraphics[scale=0.212]{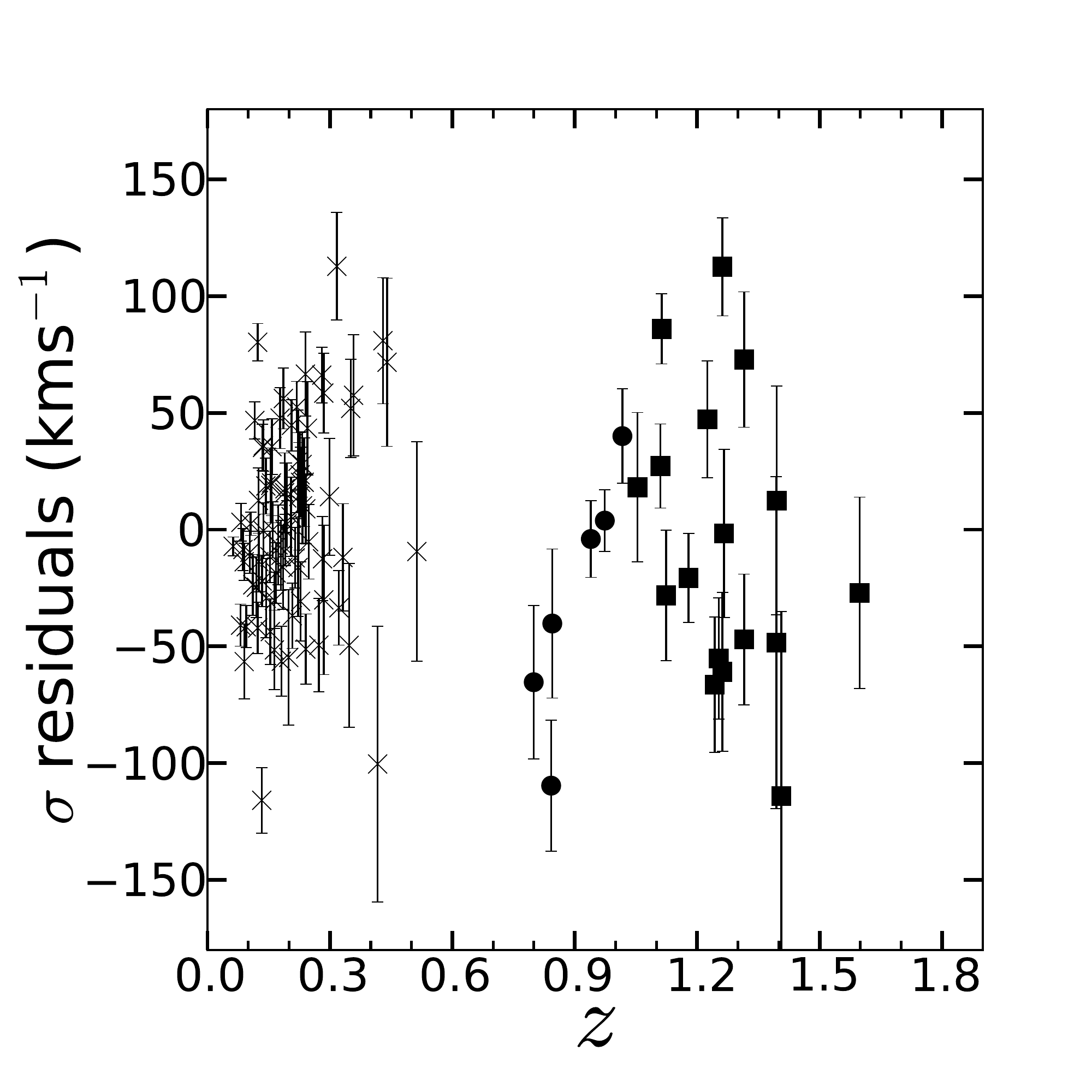}
\includegraphics[scale=0.212]{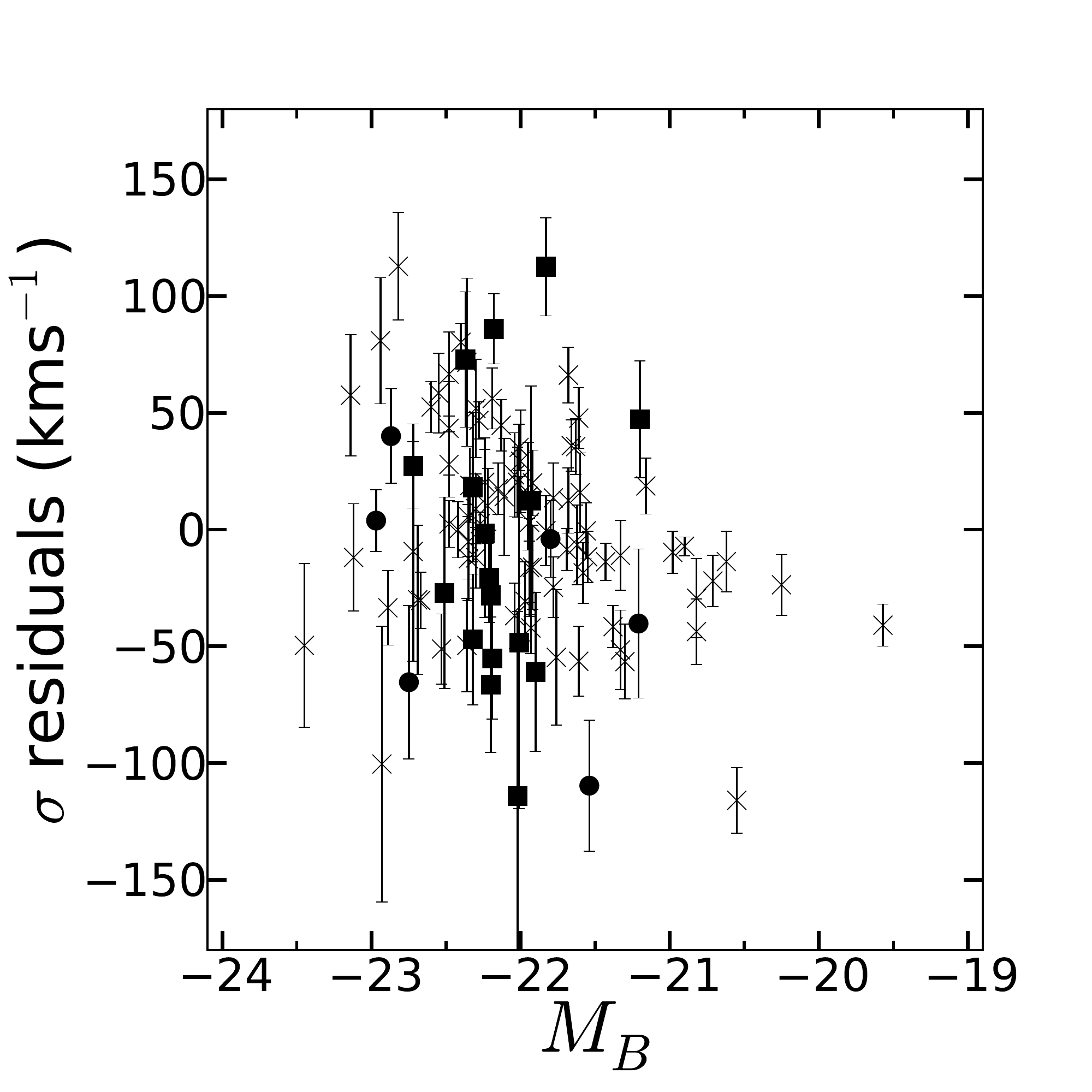}
\caption{\textbf{Left}: The Faber-Jackson Relation (FJR) that we derive (dashed) projected along the $z$-axis and the galaxies in the three samples of \citet{treu2005assembly}, \citet{auger2009sloan}, and \citet{newman2010keck}. We find no systematic biases in the FJR we derive. \textbf{Centre left}: The residuals of the velocity dispersions of the galaxies in the \citet{auger2009sloan} and \citet{newman2010keck} samples as a function of effective radius. \textbf{Centre right}: The residuals of the velocity dispersions of the galaxies in the three samples as a function of redshift. The scatter in the residuals at $z>0.6$ agrees with the scatter in the residuals at $z<0.6$ within the uncertainties. \textbf{Right}: The residuals of the velocity dispersions as a function of the $B$-band magnitude.}
\label{figure:FJR}
\end{figure*}

\indent There exists a spatial correlation between bright, high-$z$ LBGs and bright foreground objects (see Appendix \ref{appendix:spatial}). Spatial correlations between source LBGs and foreground objects suggests that magnification bias is detectable in current surveys. In order to quantify its extent, foreground objects need to be modelled as gravitational lenses.\\
\indent The key parameter determining the efficiency of an early-type galaxy as a gravitational lens is its velocity dispersion \citep{turner1984statistics}. We estimate the velocity dispersion of each galaxy in the CANDELS field from its photometry by calibrating a redshift-dependent Faber-Jackson relation \citep[][FJR]{faber1976velocity}. The FJR relates the luminosity of an object to its velocity dispersion. We include a redshift evolution term to account for the evolution of the mass-to-light ratio with increasing redshift, so the FJR can be expressed as
\begin{equation}
L_{B}=m\sigma_{\star}^{\gamma}(1+z)^{\beta},
\label{eqn:FJR1}
\end{equation}
where $L_{B}$ is the $B$-band luminosity, $\sigma_{\star}$ is the stellar velocity dispersion, and $z$ is the redshift. The FJR can be expressed linearly as
\begin{equation}
M_{B} = ax + by + c,
\label{eqn:FJR2}
\end{equation}
where $a=-2.5\gamma$, $x=\log_{10}\frac{\sigma_{\star}}{200\mathrm{kms}^{-1}}$, $b=-2.5\beta$, $y=\log_{10}(1+z)$, $c=-2.5\log_{10}(m)$ and $M_{B}=-2.5\log_{10}(L_{B})$. We calibrate the FJR using galaxies with spectroscopic redshifts and velocity dispersions from \citet{treu2005assembly}, \citet{auger2009sloan}, and \citet{newman2010keck}, which span $0<z<1.6$. We determine the values of $a$, $b$, and $c$ by minimising the $\chi^{2}$, given by
\begin{equation}
\chi^2 = \sum_{i=0}^{n}\frac{(M_{B_{i}}-ax_{i}-by_{i}-c)^{2}}{(\epsilon_{M_{B_{i}}}^{2} + a^{2}\epsilon_{x_{i}}^{2} + b^{2}\epsilon_{y_{i}}^{2} + \epsilon_{\textrm{int}}^{2})},
\label{eq2}
\end{equation}
where $\epsilon_{M_{B},x,y}$ are the uncertainties in the data and $\epsilon_{\textrm{int}}$ is the intrinsic scatter. To avoid degeneracies, we fix the slope to be $\gamma=3.9$, in line with previous studies \citep{hyde2009curvature, jonsson2010weighing}.\\
\indent We find best fit parameters of $m=2.3\pm0.2\times10^{8}$ and $\beta=0.7\pm0.3$. The errors on $m$ and $\beta$ are not independent, so the uncertainty in the inferred velocity dispersion due to their uncertainty is $\sim10$kms$^{-1}$, and will not significantly affect the inferred strongly-lensed fraction. The FJR is plotted in the left panel of Figure \ref{figure:FJR}, and the residuals are plotted as a function of effective radius, $R_{\textrm{e}}$, redshift, $z$, and $B$-band magnitude, $M_B$ are plotted in the centre left, centre right and right panels, respectively. The uncertainty in the FJR is dominated by the intrinsic scatter, which is $46$kms$^{-1}$ in the direction of velocity dispersion. There are no systematic biases in the residuals with respect to $M_{B}$, $z$ or the $R_\textrm{e}$. The scatter in the residuals for galaxies at $z>0.6$ is consistent with galaxies at $z<0.6$, with no evidence of redshift-dependent scatter in our FJR.\\
\indent The resultant FJR is consistent with $B$-band FJRs found from weak lensing analyses of type Ia supernovae presented by \citet{jonsson2010weighing, kleinheinrich2004weak, hoekstra2004properties}. As a check of our FJR, we compare it with the $i^{\star}$-band FJR (which may be less prone to dust-extinction) presented by \citet{bernardi2003early} for all objects in GOODS. We find very close agreement between $\sigma_{\star}$ as inferred from the $i$-band FJR with our $B$-band FJR. For low-redshift objects, the scatter in the residuals between the two methods is $2$kms$^{-1}$. For all objects out to $z=2$, the scatter in the residuals between the two FJRs is $9$kms$^{-1}$. This may be partially due to the \citet{bernardi2003early} FJR being calibrated at $z\sim0$, and not taking into account redshift evolution. There are no systematic biases in the residuals between these two FJRs as a function of $z$, $M_{B}$ or $R_{\textrm{e}}$.\\
\indent We compare our FJR with estimates of velocity dispersions using the stellar mass estimates of galaxies in our calibration sample using the relation between stellar mass and velocity dispersion presented by \citet{hyde2009luminosity}. We find the scatter in the residuals of velocity dispersion estimates using this method to be very similar (in fact, slightly larger) than those found using our FJR. This suggests that using stellar mass information rather than $L_B$ will not significantly reduce the scatter in our velocity dispersion estimates.\\

\section{Assessing the Strong Lensing Likelihood of Lyman-Break Galaxies}
\label{section:method}
\indent To quantify the strongly-lensed fraction of LBGs, we model every foreground object in the field as a gravitational lens. Using photometric information of all foreground objects, we ask the following question for each LBG: \textit{what is the likelihood of this LBG being gravitationally lensed with magnification $\mu>2$ given its position relative to nearby (in projection) foreground objects?} We disregard deflector-LBG pairs with a separation of $\theta_{\textrm{sep}}>5\farcs0$, which is much larger than the Einstein Radius of typical deflectors. While the choice of maximum separation of $5\farcs0$ is somewhat arbitrary, we show in Section \ref{section:lensproperties} that $5\farcs0$ is a reasonable choice. For each foreground object within $5\farcs0$ of the LBG, we use the following process:
\begin{enumerate}
\item Model the foreground object using a singular isothermal sphere (SIS) density profile,
\item Calculate the velocity dispersion that the foreground object requires for it to produce an image at the observed position of the LBG with a magnification of $\mu=2$, denoted by $\sigma_{\star,\textnormal{req}}$. For an SIS, $\mu=2$ marks the beginning of the strong lensing regime. The required velocity dispersion depends on the LBG-deflector separation, LBG redshift and deflector redshift,
\item Calculate the likelihood that the foreground object has a velocity dispersion greater than or equal to $\sigma_{\star,\textnormal{req}}$. This is the likelihood of strong lensing for that deflector-LBG pair,
\item Weight the likelihood of lensing by the inverse of the detection completeness at the separation between the LBG and the nearby foreground object.
\end{enumerate}
The final step accounts for reduced sensitivity to faint LBGs nearby bright foregrounds. We explain this process further in Section \ref{subsection:sv}.\\ 
\indent To calculate $\sigma_{\star,\textnormal{req}}$ we find the Einstein Radius, $\theta_{\textrm{ER}}$, required for $\mu=2$ using the expression for the magnification of the image in an observed configuration,
\begin{equation}
\mu = \frac{|\theta_{\textrm{sep}}|}{|\theta_{\textrm{sep}}|-\theta_{\mathrm{ER}}},
\end{equation}
where $\mu$ is the magnification, and $\theta_{\textrm{sep}}$ is the observed separation between the source image and the deflector. We can then find the velocity dispersion corresponding to $\mu=2$ using the expression for the Einstein Radius of an SIS,
\begin{equation}
\theta_{\mathrm{ER}}=4\pi(\frac{\sigma_{\star}}{c})^{2}\frac{D_{LS}}{D_{S}},
\end{equation}
where $\sigma_{\star}$ is the stellar velocity dispersion, $D_{S}$ is the angular diameter distance to the source, and $D_{LS}$ is the angular diameter distance from the lens to the source.\\
\indent For each LBG-foreground object pair, the likelihood of strong lensing of the LBG by the deflector is equal to the likelihood that the deflector has a velocity dispersion above $\sigma_{\star\textnormal{,req}}$, which is given by
\begin{equation}
\mathscr{L}=\frac{1}{2} \textnormal{erfc} \Big( \frac{\sigma_{\star\textnormal{,req}}-\sigma_{\star\textnormal{,inf}}}{\sqrt 2 \epsilon_{\textnormal{FJR}}} \Big),
\label{eqn:l}
\end{equation}
where $\sigma_{\star\textnormal{,inf}}$ is the velocity dispersion inferred from photometry (using the FJR), and $\epsilon_{\textnormal{FJR}}$ is the intrinsic scatter in the velocity dispersion of the FJR.\\
\indent In the event that there are multiple potential deflectors within $5\farcs0$ of the source, we treat them independently and calculate the probability that at least one is lensing the source by $\mu\geq2$. For $n$ deflectors, this is
\begin{equation}
\mathscr{L}=1-\prod_{j=1}^{n}(1-\mathscr{L}_{j}).
\end{equation}
\indent We show a subset of the sample consisting of some of the highest-likelihood lenses in Figure \ref{figure:lenses}. We describe these systems further in Section \ref{section:examples}.\\

\subsection{Accounting for Sensitivity Variations}
\label{subsection:sv}
\indent Faint LBG samples have a reduced completeness compared to bright ones. This alone would not affect our inference of the lensed fraction, because the completeness would change the numerator and the denominator by the same factor at fixed magnitude. However, we note that there may be a further reduced sensitivity to detecting faint LBGs around bright objects, which will affect potentially lensed LBGs differently to those isolated in the field. This effect could cause our measured strongly lensed fraction to be artificially low.\\
\indent We weight all LBGs that appear close in projection to bright foreground objects by the inverse of their relative detection probability in order to account for reduced sensitivity around foreground objects. To do this, we run completeness simulations around all foreground objects which are either,
\begin{enumerate}
\item Assessed as having a greater than $1\%$ chance of lensing a nearby LBG, or,
\item Brighter than $m_r=24$ mag and within $2\farcs5$ of an LBG.
\end{enumerate}
\indent We run source recovery simulations in order to determine completeness as a function of radius around each foreground meeting either of the above criteria. The source recovery simulations are run for LBGs\footnote{The artifical sources in the recovery simulations are extended, and have sizes typical of LBGs at the appropriate redshift. It should be noted that strong lensing may cause the sources to appear more extended, which will affect their completeness. This effect is expected to be small, but is a slight limitation of this analysis.} at the redshift and of the magnitude corresponding to that of the nearby LBG. The completeness of a source LBG, $s_c$, becomes unaffected by typical foregrounds at a separation of around $1\farcs5-2\farcs0$. The weight, $w_c$, we apply to each LBG is defined as the inverse of the completeness, $w_c\equiv 1/s_c$. We apply a maximum weighting of $w_c=10$ to any LBG.\\
\indent We find that weighting the lensed fraction in this way has a minimal effect on the bright-end of the observed lensed fraction. However, the relative completeness of faint galaxies near to bright foreground galaxies is, as expected, lower than for the brighter LBGs..\\

\subsection{Uncertainty Checks}
\label{subsection:uncertainties}
\indent The typical uncertainty in the photometric redshifts of foreground sources derived in the 3D-HST catalogues are $\Delta z\sim0.1-0.2$, so the uncertainty in the magnification is dominated by the intrinsic uncertainty in the FJR. Furthermore, the uncertainty in $M_{B}$ due to photometric redshift errors (via the distance modulus) is partially self-regulating as the inferred velocity dispersion is a decaying function of redshift, while inferred rest-frame luminosity is an increasing function. We find that of the $40$ $z\sim7$ LBGs that have a likelihood of lensing of $\geq 10\%$, the deflector of only one has a photometric redshift with less than an $80\%$ chance of residing within $\Delta z=0.2$. None of the deflectors of $z\sim4-6$ LBGs with a likelihood of lensing of $\geq 10\%$ have photometric redshifts with less than an $80\%$ chance of residing within $\Delta z=0.2$. The uncertainty in source redshift ($\Delta z\sim0.35$) is negligible as the angular diameter distance is a relatively flat function at high-redshift.\\
\indent There is a possible Eddington bias stemming from uncertainties in the photometry and the shape of the LF at $z\sim1$, which could bias the inference of $\sigma_{\star}$ from the $B$-band luminosity. We find that $\Psi(L\pm\delta L)$ only varies by $2-5\%$ from $\Psi(L)$ for galaxies brighter than $\sim M_{\star}$ at $z=1$ in the field. Therefore the number density of bright galaxies does not change significantly within the photometric uncertainties, and the Eddington bias is negligible.\\
\begin{figure*}
\begin{center}
\includegraphics[trim=23 45 114 110, clip, scale=0.115]{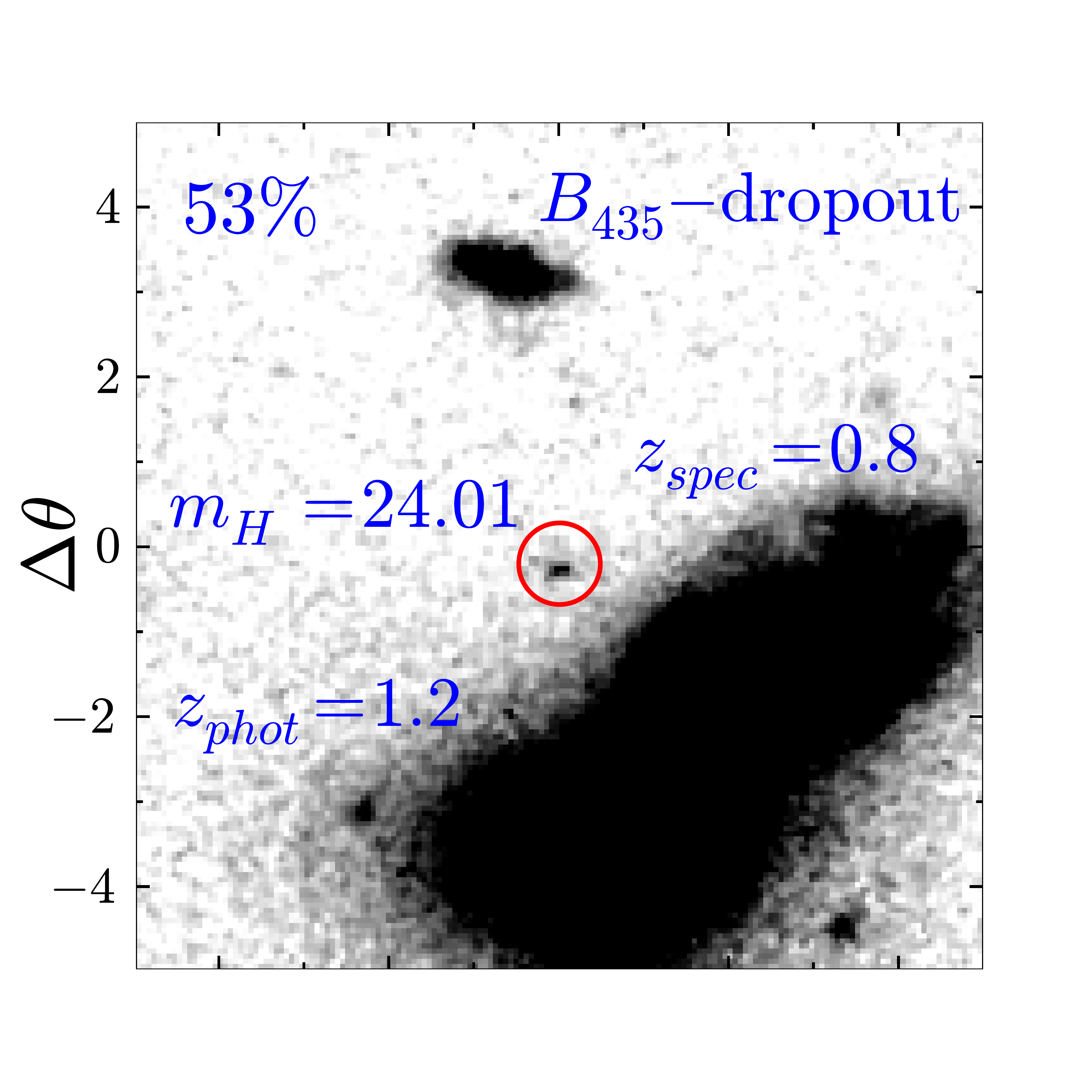}
\includegraphics[trim=23 45 114 110, clip, scale=0.115]{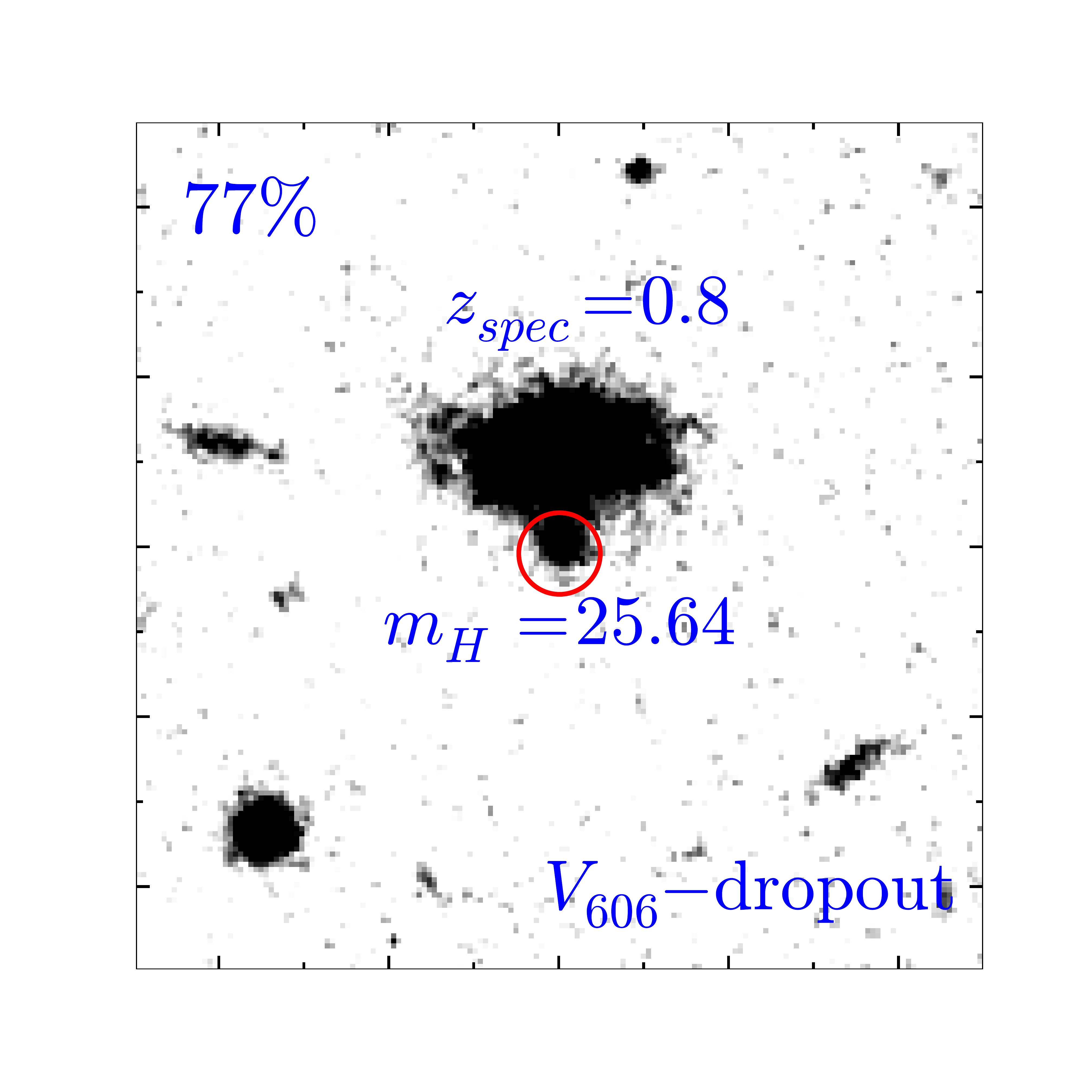}
\includegraphics[trim=23 45 114 110, clip, scale=0.115]{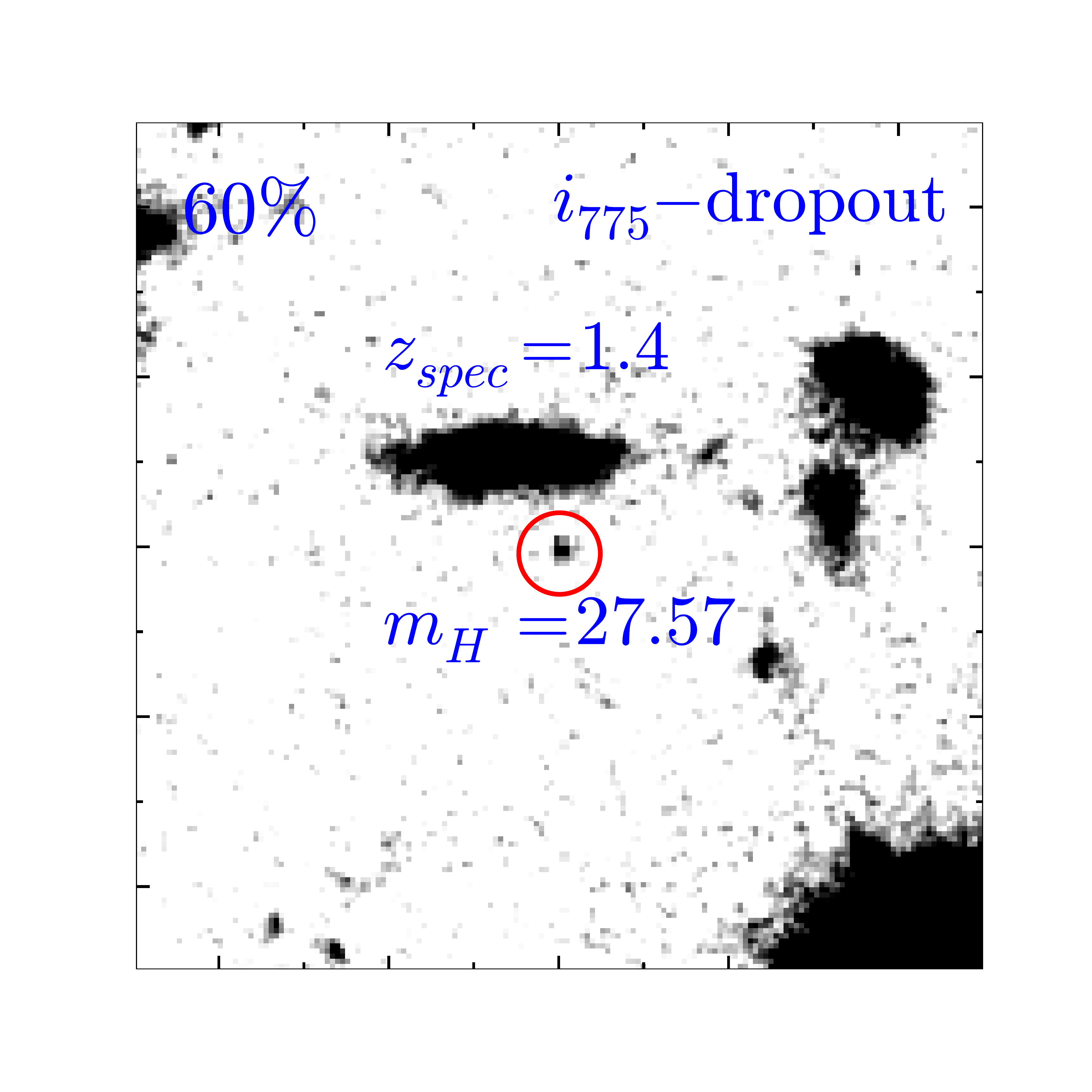}
\includegraphics[trim=23 45 60  110, clip, scale=0.115]{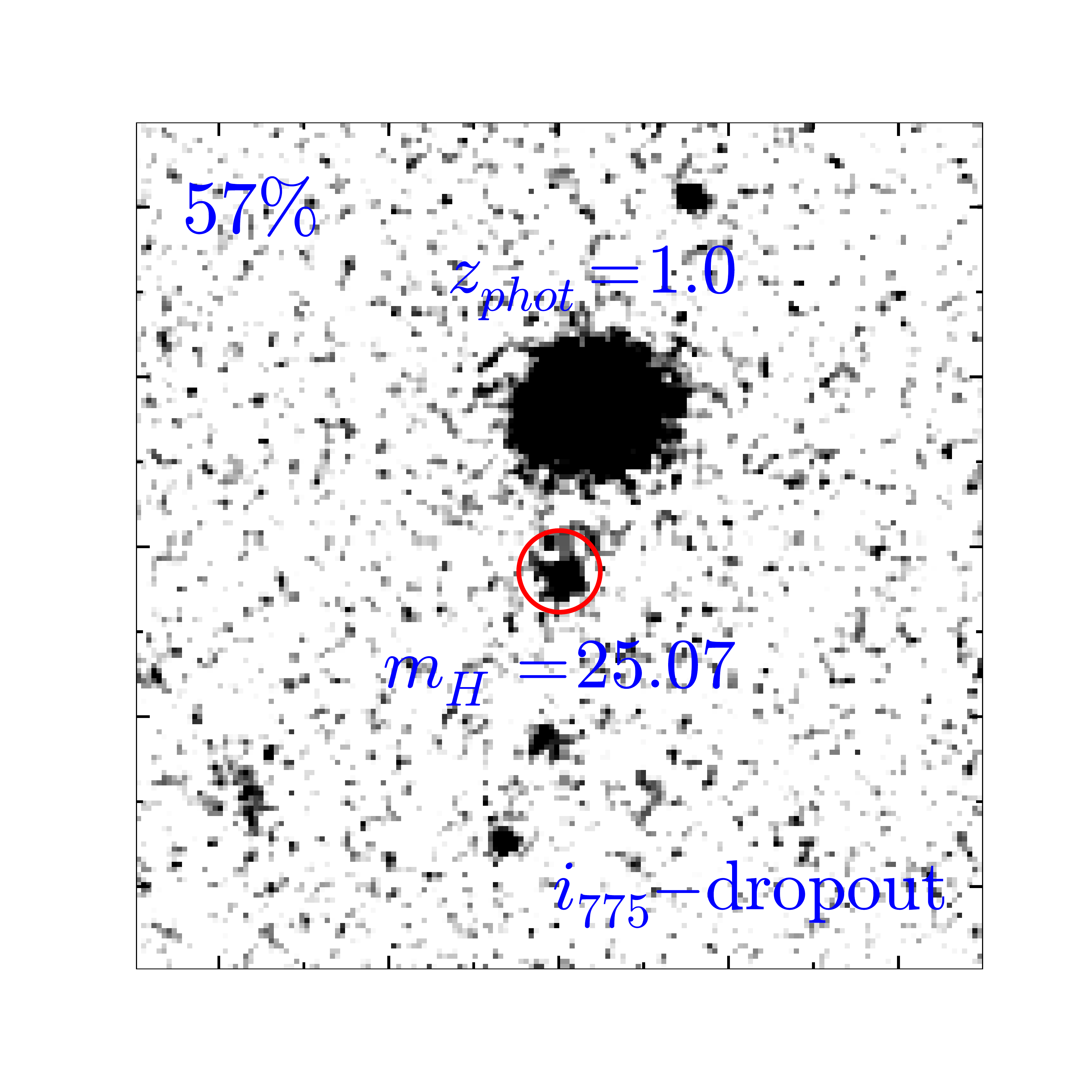}\\
\includegraphics[trim=23 5  114 110, clip, scale=0.115]{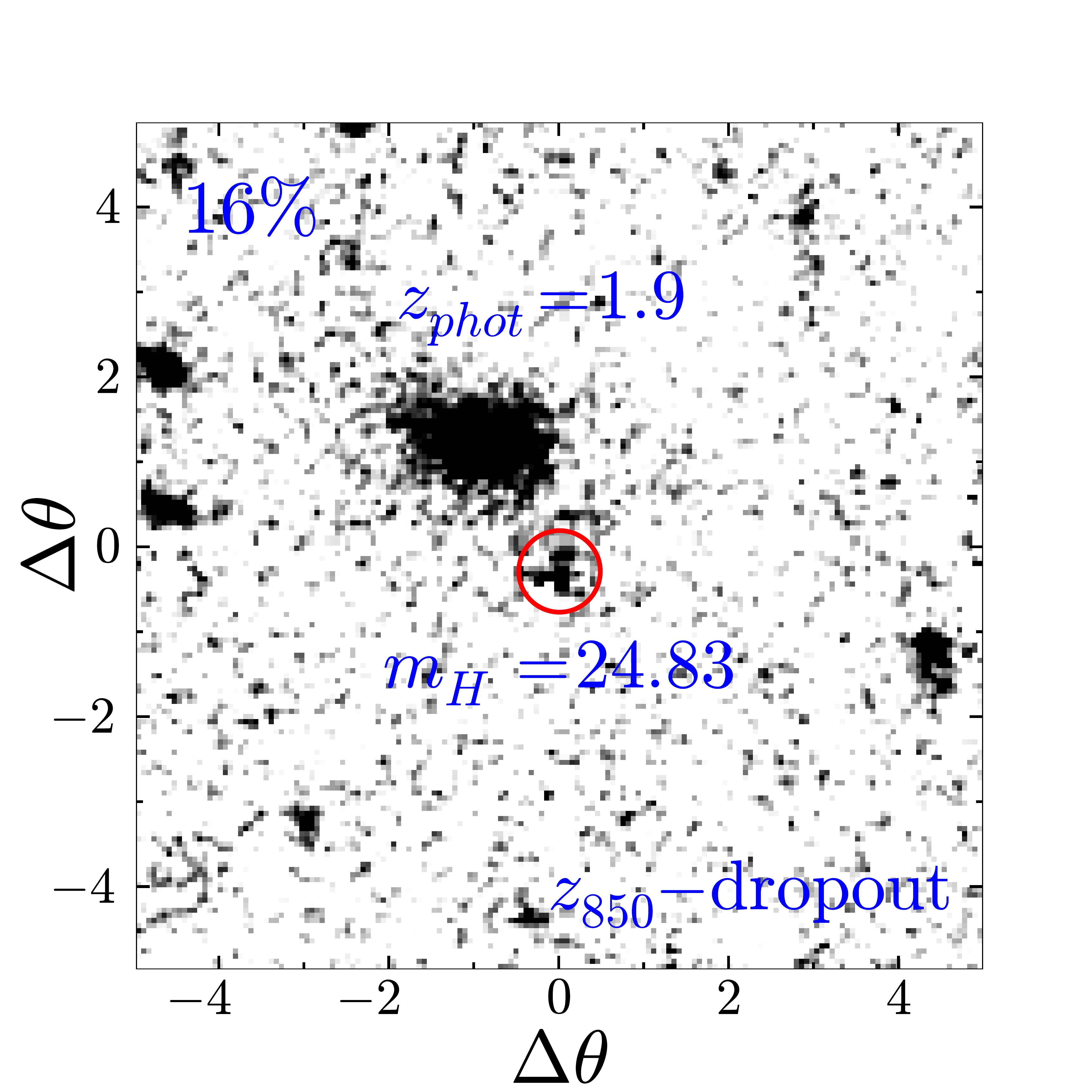}
\includegraphics[trim=23 5  114 110, clip, scale=0.115]{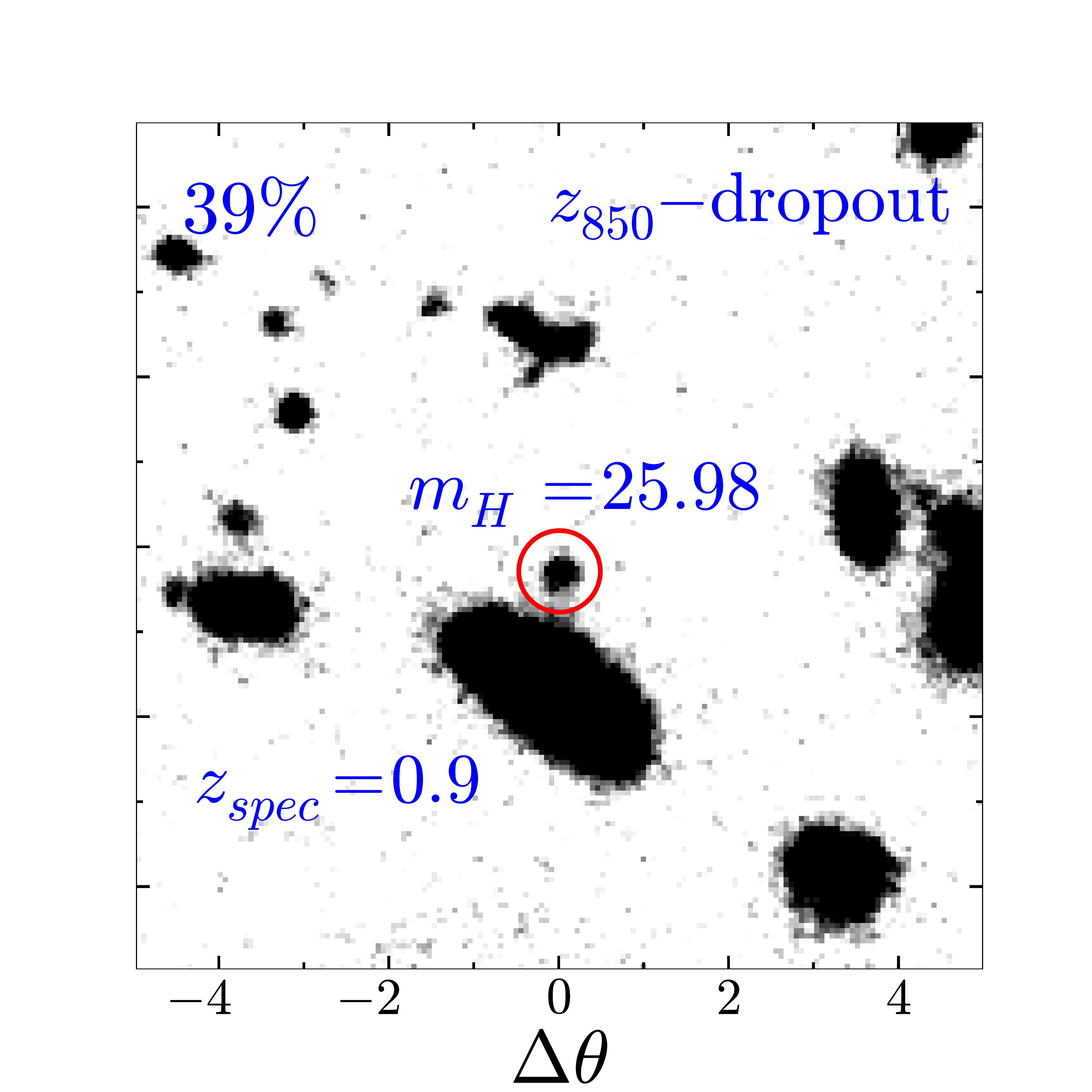}
\includegraphics[trim=23 5  114 110, clip, scale=0.115]{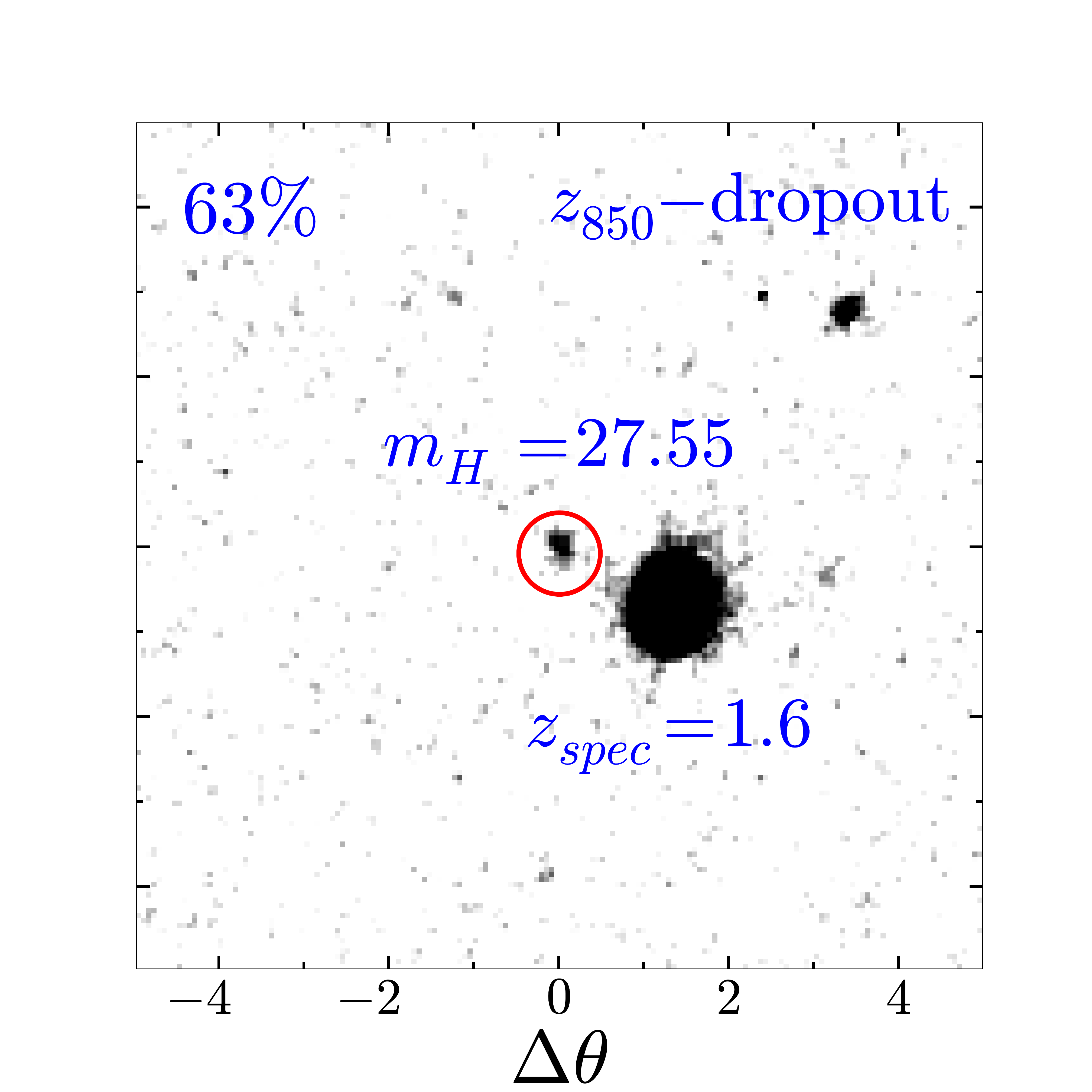}
\includegraphics[trim=23 5  60  110, clip, scale=0.115]{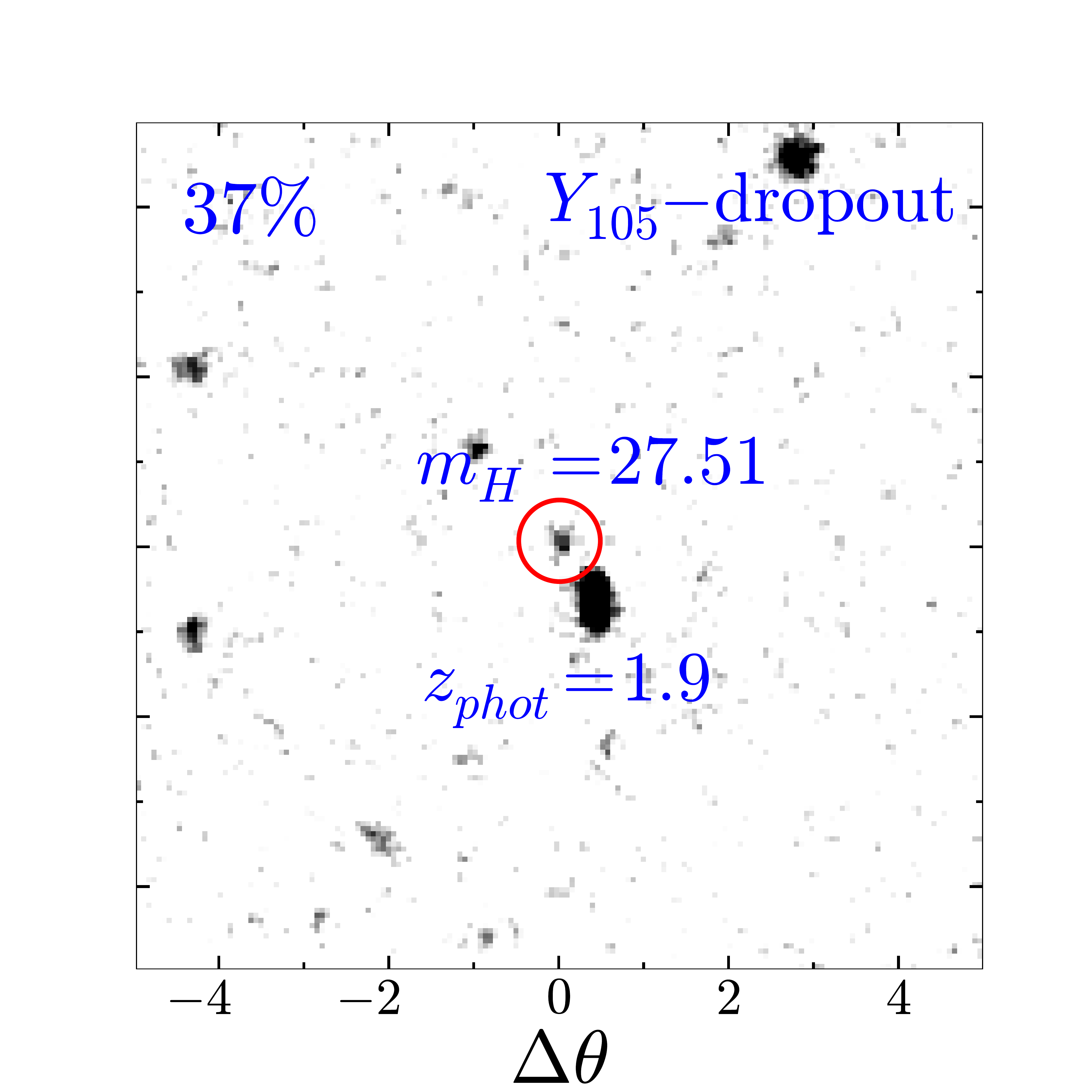}
\caption{Examples of possibly-lensed LBGs in the four samples. All cutouts are of the $J_{125}$ images, are $10\farcs0\times10\farcs0$ and are shown on the same contrast scale (except for the top, left cutout, which contains two very bright foreground galaxies). The LBGs are circled in red and the deflectors are labelled by their spectroscopic/photometric redshifts. Each LBG is labelled with its $H_{160}$ magnitude, and its likelihood of being strongly lensed (top left corner). The LBG shown in the bottom, left panel is the brightest LBG in the $z\sim7$ sample.}
\label{figure:lenses}
\end{center}
\end{figure*}
\indent We note that a limitation of this analysis is that it assumes all deflectors to be singular isothermal spheres. Singular isothermal ellipsoids (SIEs) may be a more realistic parametrization of potential deflectors \citep[see][for ellipsoidal density parametrizations]{keeton2001catalog}. However, SIEs do complicate the calculation significantly \citep{kormann1994isothermal, huterer2005effects}. The median ellipticity for all objects in the CANDELS fields is $\overline{\epsilon}=0.21$, with $83\%$ having an ellipticity of $\epsilon<0.4$. In the case of low deflector ellipticity ($\epsilon\lesssim0.2$), the change in the magnification estimate is $\approx10\%$ along both the major ($+10\%$) and minor ($-10\%$) axes. For larger ellipticities ($\epsilon\simeq0.4$), the magnification estimate becomes $\approx20\%$ lower for an image located along the minor axis, and a factor of two higher for images along the major axis. Using an elliptical deflector model for the system shown in the bottom, centre-left panel of Figure \ref{figure:lenses}, which includes a deflector with large ellipticity ($\epsilon=0.48$), we estimate the magnification to be $\mu\simeq1.6$, as compared with the SIS estimation of $\mu\simeq1.8$. The LBG in the bottom, right panel of Figure \ref{figure:lenses}, which is near a deflector with ellipticity $\epsilon=0.33$, has a magnification of $\mu\simeq1.6$ in the SIS model, which becomes $\mu\simeq1.9$ using an ellipsoidal model. \\
\indent Similarly, the lensing cross section (the strong lensing area in the image plane owing to a deflector) of an SIE is the same as the optical depth of a SIS with a higher-order term \citep{kormann1994isothermal}. The area of sky covered by the Einstein Radius of an SIE is only $\approx 5\%$ larger than the area of sky covered by an SIS for reasonable ellipticities ($\epsilon\lesssim0.4$). Therefore our calculations of the optical depth in Section \ref{section:magbias} are not significantly affected by the SIS assumption. These calculations are consistent with previous studies of the effect of ellipticity on the strong lensing optical depth and magnification, such as \citet{huterer2005effects} who noted that aside from image multiplicities, introducing shear and ellipticity has surprisingly little effect. Hence, using an SIE deflector will not qualitatively change either our strongly lensed fraction or magnification bias results.\\

\section{The Strongly Lensed Fraction}
\label{section:results}
\begin{figure*}
\includegraphics[scale=0.23]{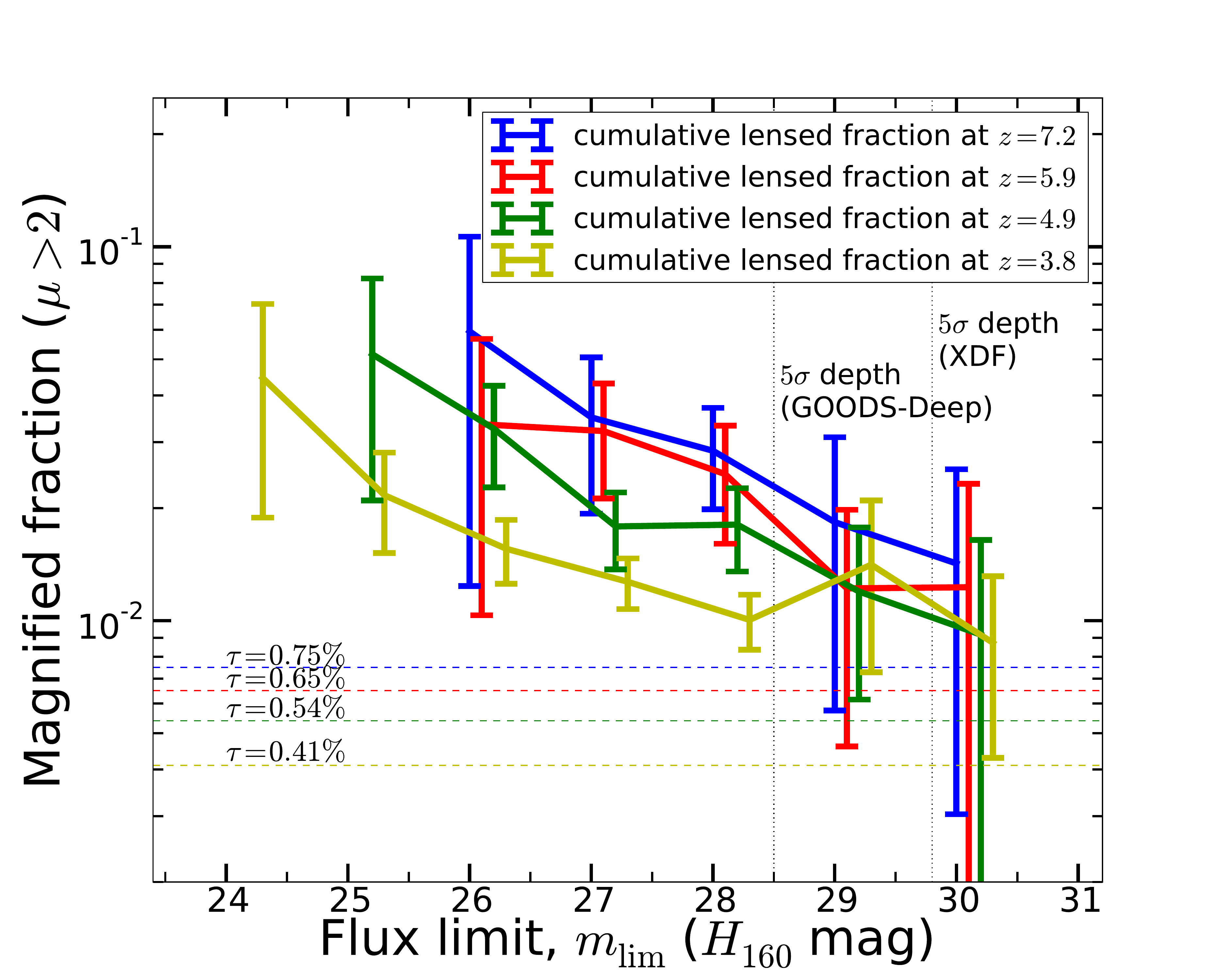}
\includegraphics[scale=0.23]{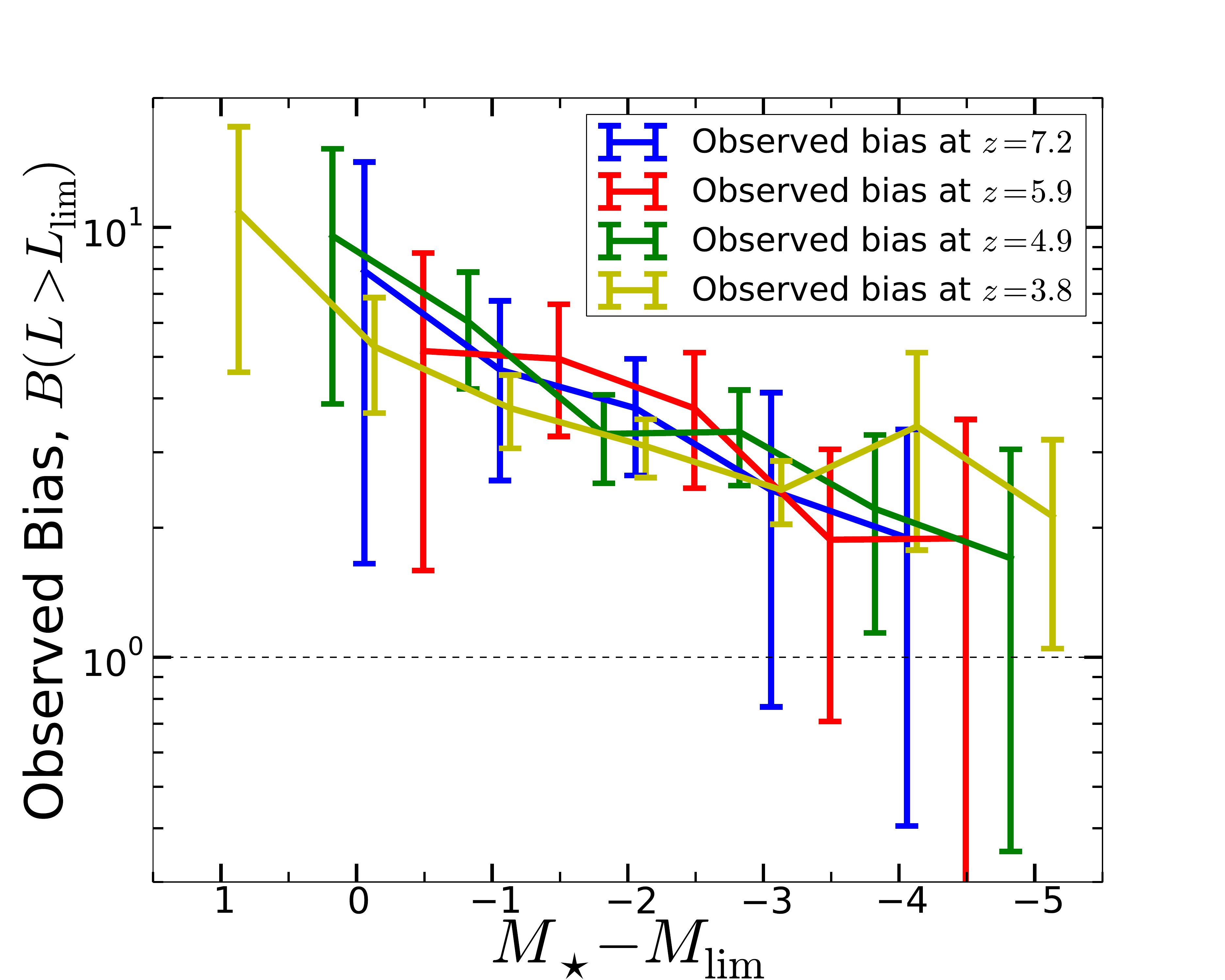}
\caption{\textbf{Left:} The lensed fraction of background LBGs as a function of flux limit for the $z_{850}$ and $Y_{105}$-dropout samples (blue), $i_{775}$-dropouts (red, offset by $m+0.1$), $V_{606}$-dropouts (green, offset by $m+0.2$) and $B_{435}$-dropouts (yellow, offset by $m+0.3$). The observed lensed fraction decreases monotonically with decreasing redshift for flux limits of $m_{\textrm{lim}}=26$, $27$, $28$ \& $30$. The analytic strong lensing optical depths, $\tau$, (strongly-lensed fraction of random lines of sight) for each source redshift are plotted as dashed lines in the same colours as the observed lensed fractions (see Section \ref{section:magbias}). \textbf{Right:} The observed magnification bias at each redshift overlaid as a function of $M_{\star}-M_{\textrm{lim}}$ (see Section \ref{section:magbias}). The $B_{435}$ sample contains the brightest measurements with respect to $M_{\star}$ ($1$ mag brighter), followed by the $V_{606}$ sample, the $z_{850}$ \& $Y_{105}$ sample, and the $i_{775}$ sample. The bias is defined as the ratio of the solid and dashed lines in the left panel. We show the bias at each redshift individually in Figure \ref{figure:magbias}. For an LF without strong evolution in $\alpha$, which is approximately observed from $4<z<7$, the bias is not expected to evolve. A roughly constant bias is observed at all values of $M_{\star}-M_{\textrm{lim}}$ for the four independent LBG samples from $4<z<7$.}
\label{figure:magfrac}
\end{figure*}

\indent The method of prescribing a likelihood of strong lensing described in Section \ref{section:method} was applied to each LBG in the samples at $z\sim4$, $z\sim5$, $z\sim6$ and $z\sim7-8$. The $455$ $z_{850}$ dropouts and $155$ $Y_{105}$ dropouts were combined to create a statistically significant sample with a mean redshift of $\overline{z}=7.2$. The number of strongly-lensed LBGs brighter than a theoretical survey limit, $m_{\textrm{lim}}$, is assessed for each of the samples. For the $i_{775}$ and $z_{850}$ \& $Y_{105}$ samples, we assess the lensed fraction brighter than $m_{\textrm{lim}}=26,27,28,29$ \& $30$ mag in $H_{160}$. We include $m_{\textrm{lim}}=25$ mag for the $V_{606}$ sample and $m_{\textrm{lim}}=24$ \& $25$ for the $B_{435}$ sample, as $M_{\star}$ appears brighter for these samples. The strongly lensed fraction is not affected by the differing depth of the CANDELS fields and the XDF.\\
\indent The cumulative lensed fraction at each of these flux limits is the ratio of the expected number of strongly lensed LBGs (the sum of all lens likelihoods) brighter than the flux limit and the total number of LBGs appearing brighter than the flux limit. However, the cumulative lensed fraction depends on the total completeness of the combined sample. To account for incompleteness in number counts, we use the LF of \citet{bouwens2014uv}. The cumulative lensed fraction at $z\sim4$, $z\sim5$, $z\sim6$ and $z\sim7$ is shown in the left panel of Figure \ref{figure:magfrac}. The right panel of Figure \ref{figure:magfrac} shows the observed magnification bias (see Section \ref{section:magbias}). When inferring properties of the luminosity function (Sections \ref{subsection:cutoff} \& \ref{subsection:lfparams}) we use the observed lensed fraction in each bin without LF-correction so as to not presuppose the nature of the LF.\\
\indent The trend to a larger fraction of strongly-lensed galaxies for brighter flux limits is reasonably smooth, monotonic and observed in each of the four independent samples. The amplitude at all flux limits steadily increases from $z\sim4$ to $z\sim7$ (although the error bars are large), which is expected as the faint-end slope steepens and the strong lensing optical depth increases at higher redshift. The excess probability of gravitational lensing of bright galaxies is detected at high-significance in each of the samples. The lensed fraction of LBGs brighter than $m_{H_{160}}=26$ is $\sim6\%$ at $z\sim7$ and $\sim3.5\%$ at $z\sim6$, although the uncertainty is large due to the rarity of bright objects at high-redshift. At $z\sim5$, the lensed fraction at the same flux limit is $\sim3.5\%$ and at $z\sim4$ the lensed fraction is $\sim1.5\%$.\\
\indent We also assess the lensed fraction at brighter flux limits for the $z\sim4$ and $z\sim5$ samples. We find that the lensed fraction continues to rise, as expected. At $z\sim5$, $\sim5\%$ of LBGs brighter than $m=25$ are strongly lensed, and at $z\sim4$, $\sim4.5\%$ of LBGs brighter than $m=24$ are lensed.\\
\indent The errors are calculated using bootstrap resampling. The bootstrap sample is drawn from the entire sample with replacement $N=10^{4}$ times. Each time, each LBG is considered either ``lensed'' or ``not lensed'' randomly according to its likelihood of having been lensed. The lens fraction is recalculated for all limiting fluxes. The error bars represent the $1\sigma$ limits of the resultant distributions.\\

\subsection{Examples of Likely Lensed Systems}
\label{section:examples}
\indent We present an illustrative sample of some likely-lensed candidates in the surveys in Figure \ref{figure:lenses}. Cases from our highest-$z$ sample are emphasised because they are of the most interest, and have the most importance to future surveys. We note that the three brightest $z$ and $Y$-dropouts in the entire sample are each deemed to have a likelihood of lensing of $>10\%$. The brightest LBG in the $z\sim7$ sample is shown in the bottom left panel of Figure \ref{figure:lenses}. All cutouts are shown at the same contrast scale, except for the $z\sim4$ lens candidate (top left), which is in proximity to two very bright foreground galaxies, both with $M_{B}\sim-23.5$, one of which is spectroscopically confirmed at $z=0.8$. All cutouts are $10\farcs0$ on each side. In each case, the deflector candidate is labelled with its spectroscopic or photometric redshift and the LBG is labelled with its $H_{160}$ magnitude.\\
\indent The cutouts highlight the difficulty in locating secondary images in the event the LBG has been strongly lensed. A secondary image will appear closer to the foreground galaxy than the primary (circled) image, and is likely to also be appear much fainter than the primary image.\\

\subsection{Deflector Properties}
\label{section:lensproperties}
\begin{figure*}
\begin{center}
\includegraphics[trim=0 0 55 0, clip, scale=0.22]{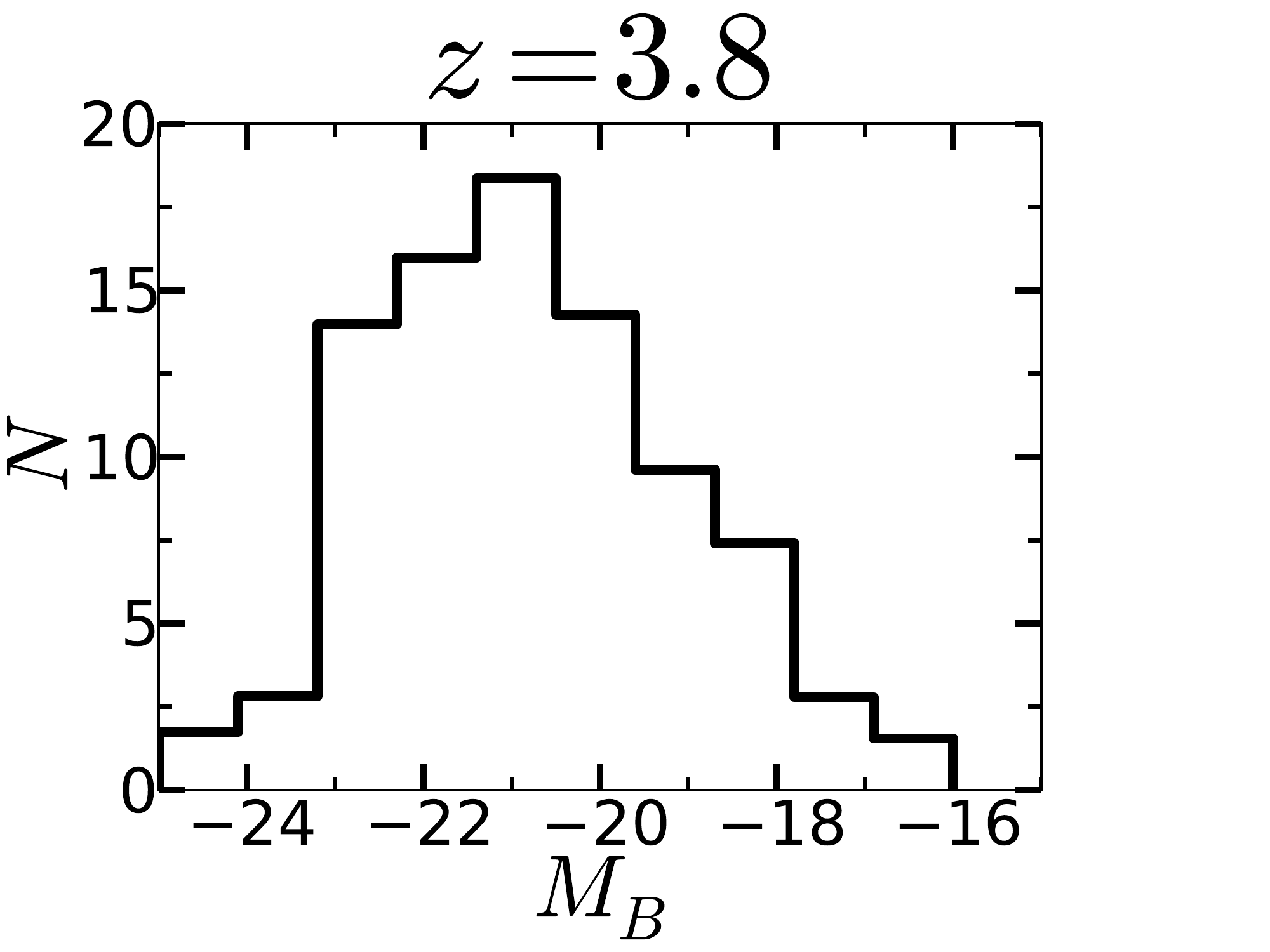}
\includegraphics[trim=0 0 55 0, clip, scale=0.22]{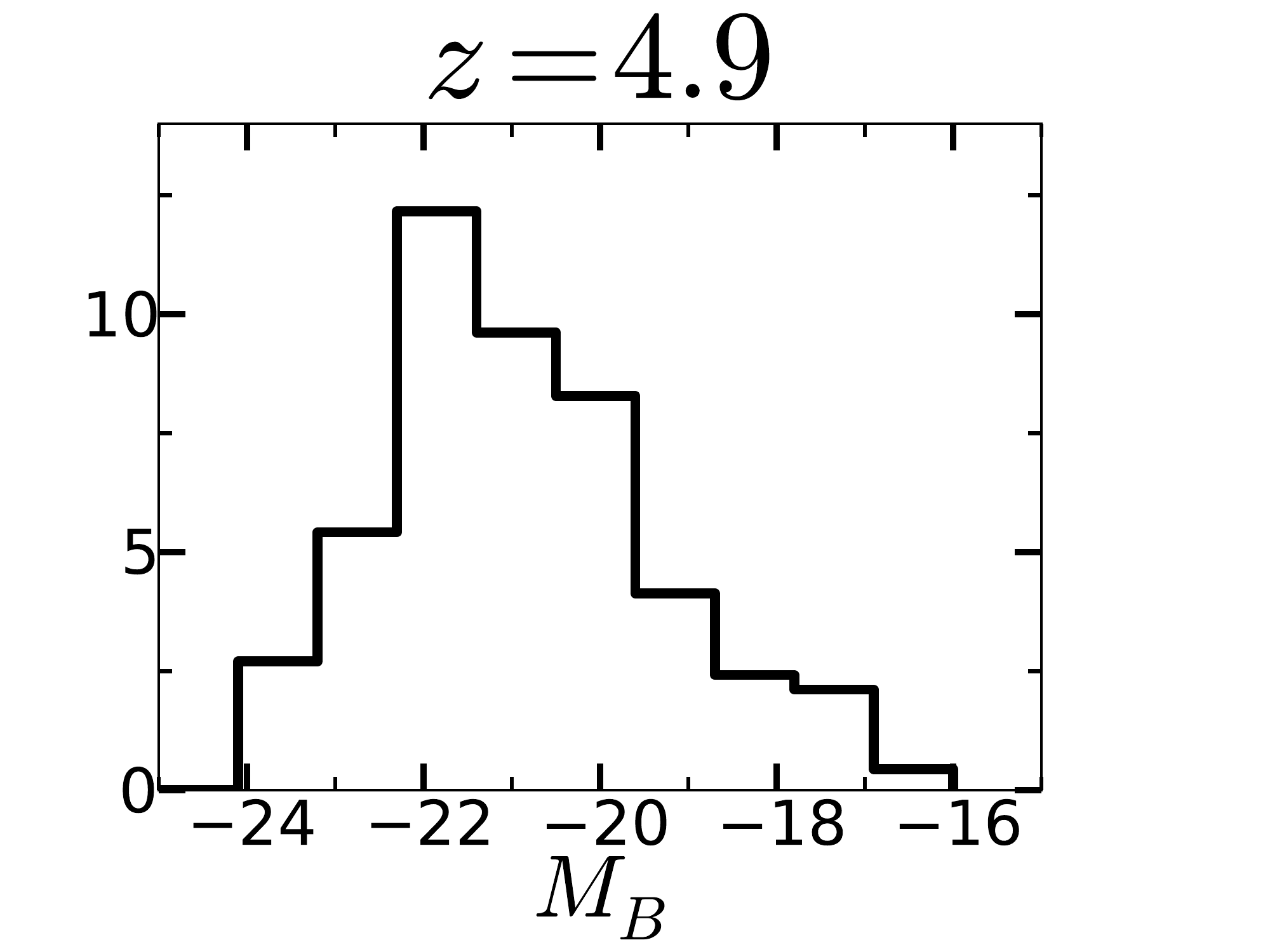}
\includegraphics[trim=0 0 55 0, clip, scale=0.22]{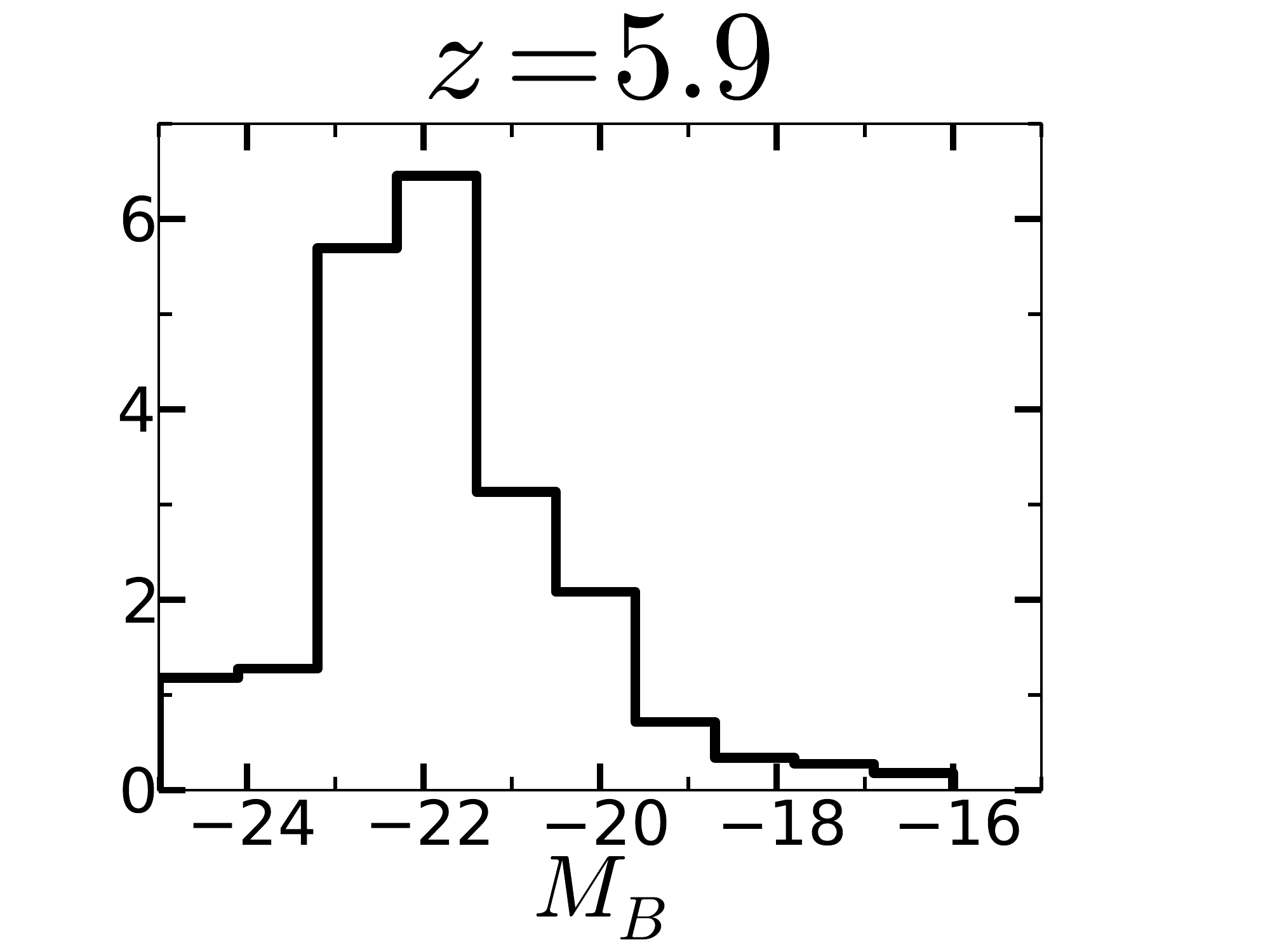}
\includegraphics[trim=0 0 10 0, clip, scale=0.22]{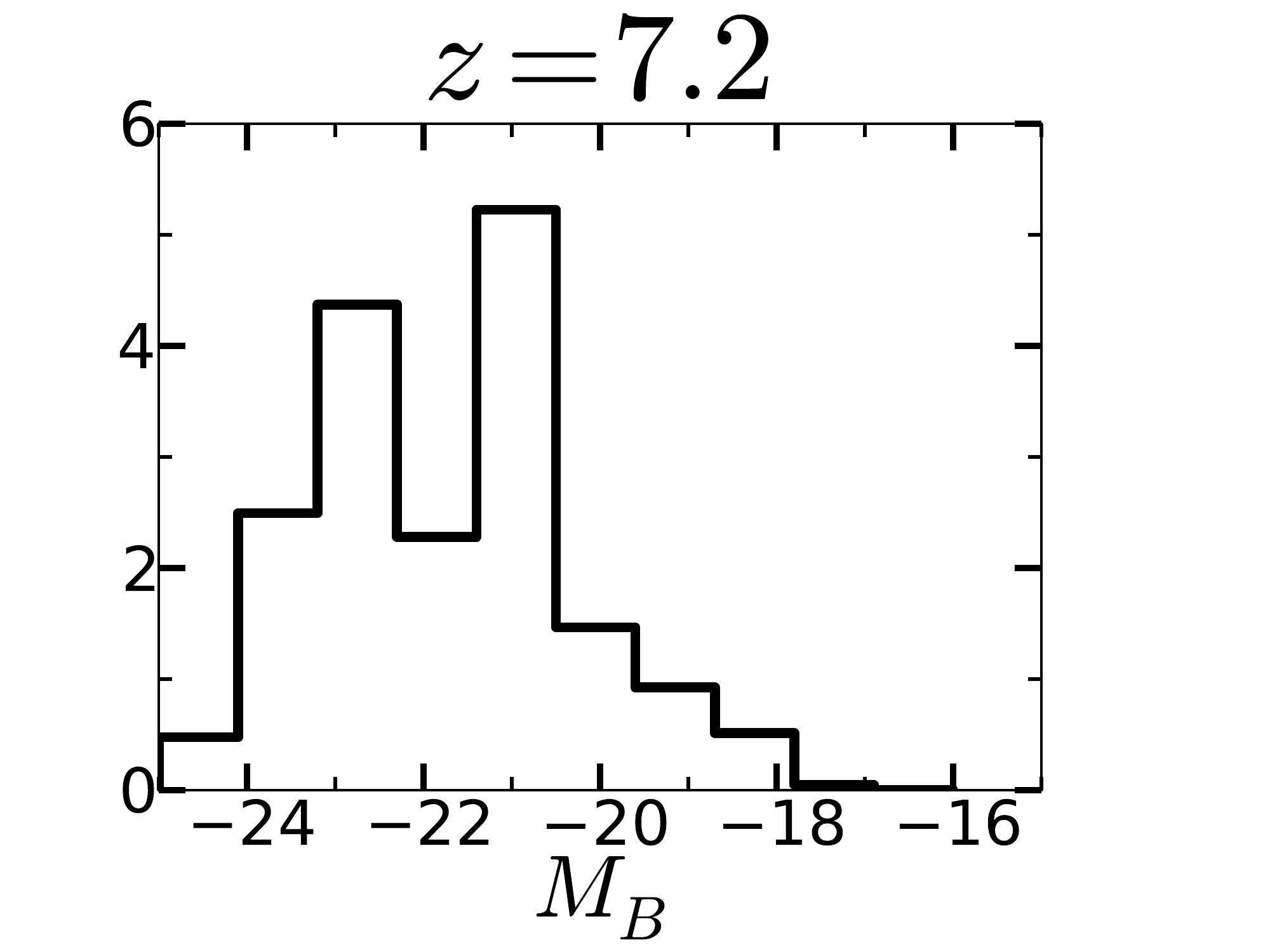}\\
\includegraphics[trim=0 10 55 0, clip, scale=0.22]{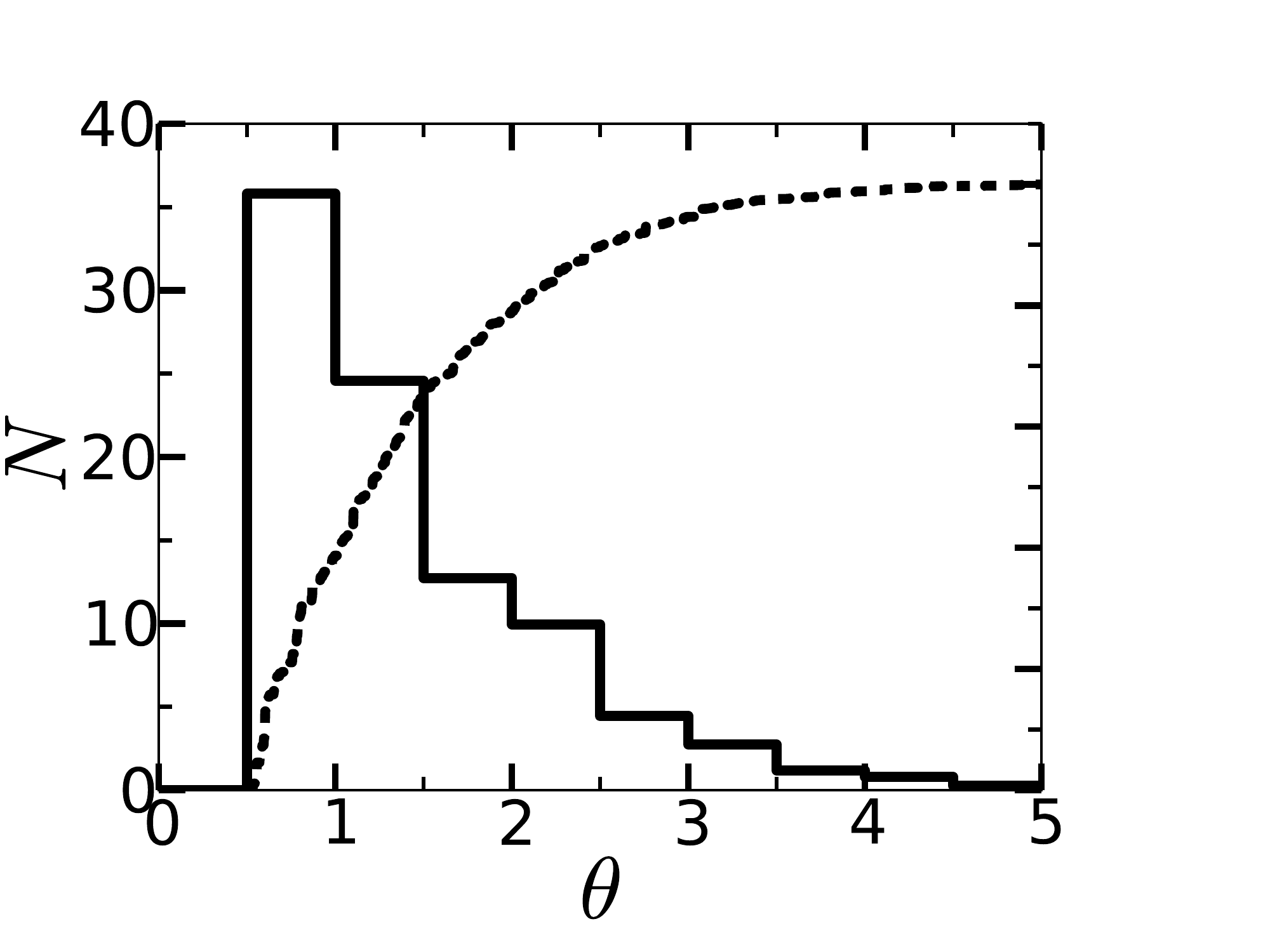}
\includegraphics[trim=0 10 55 0, clip, scale=0.22]{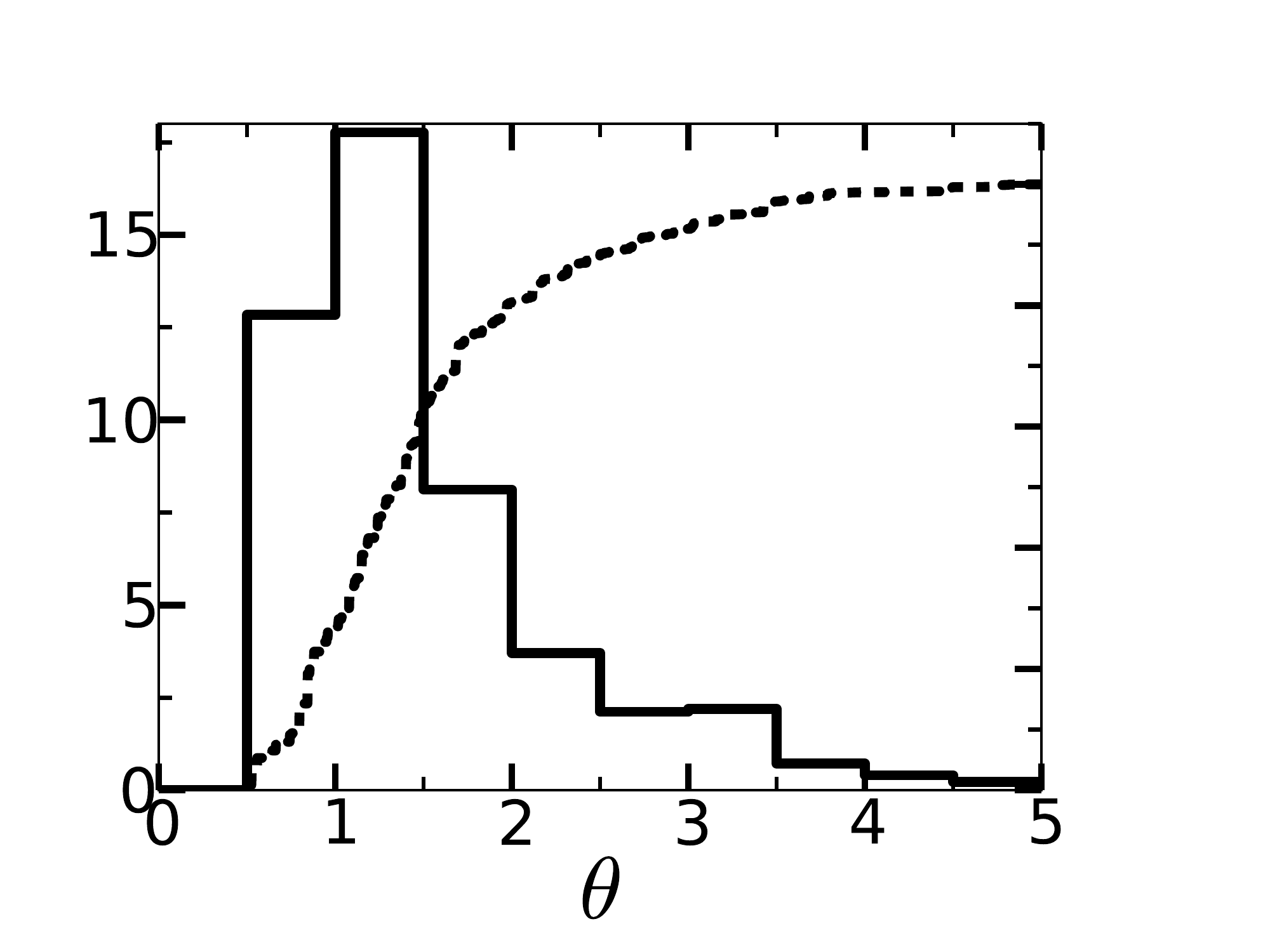}
\includegraphics[trim=0 10 55 0, clip, scale=0.22]{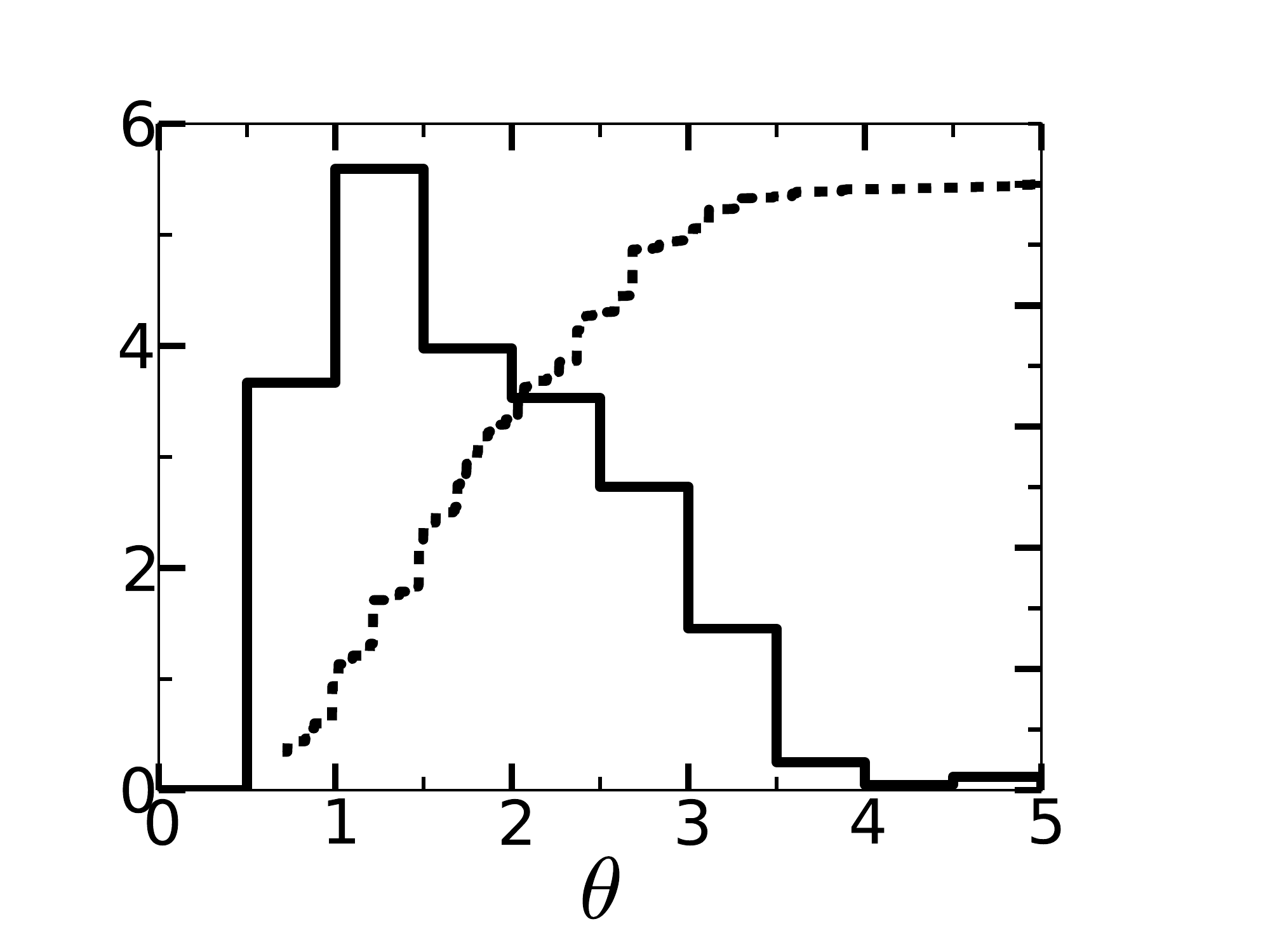}
\includegraphics[trim=0 10 10 0, clip, scale=0.22]{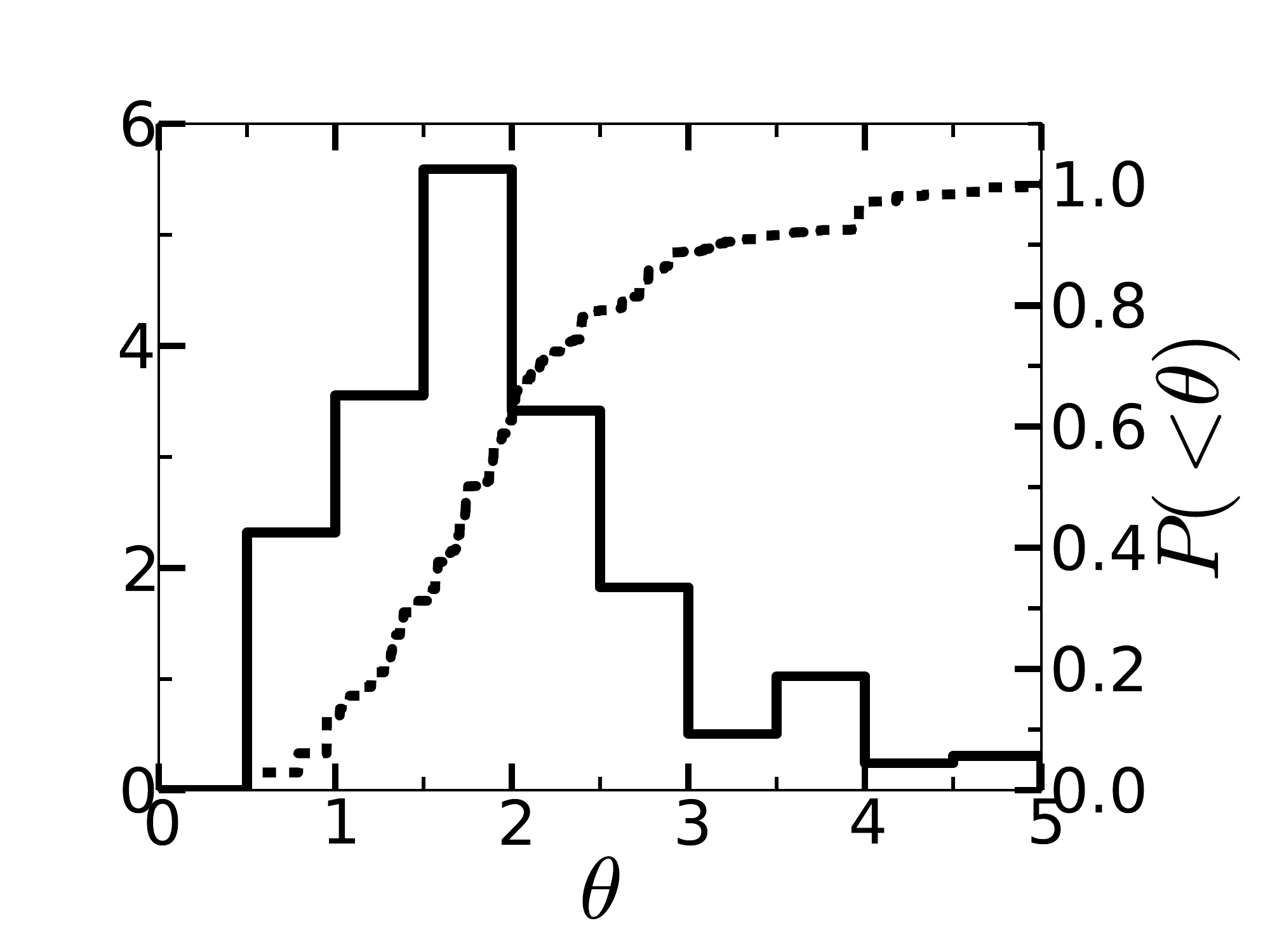}\\
\includegraphics[trim=0 10 55 0, clip, scale=0.22]{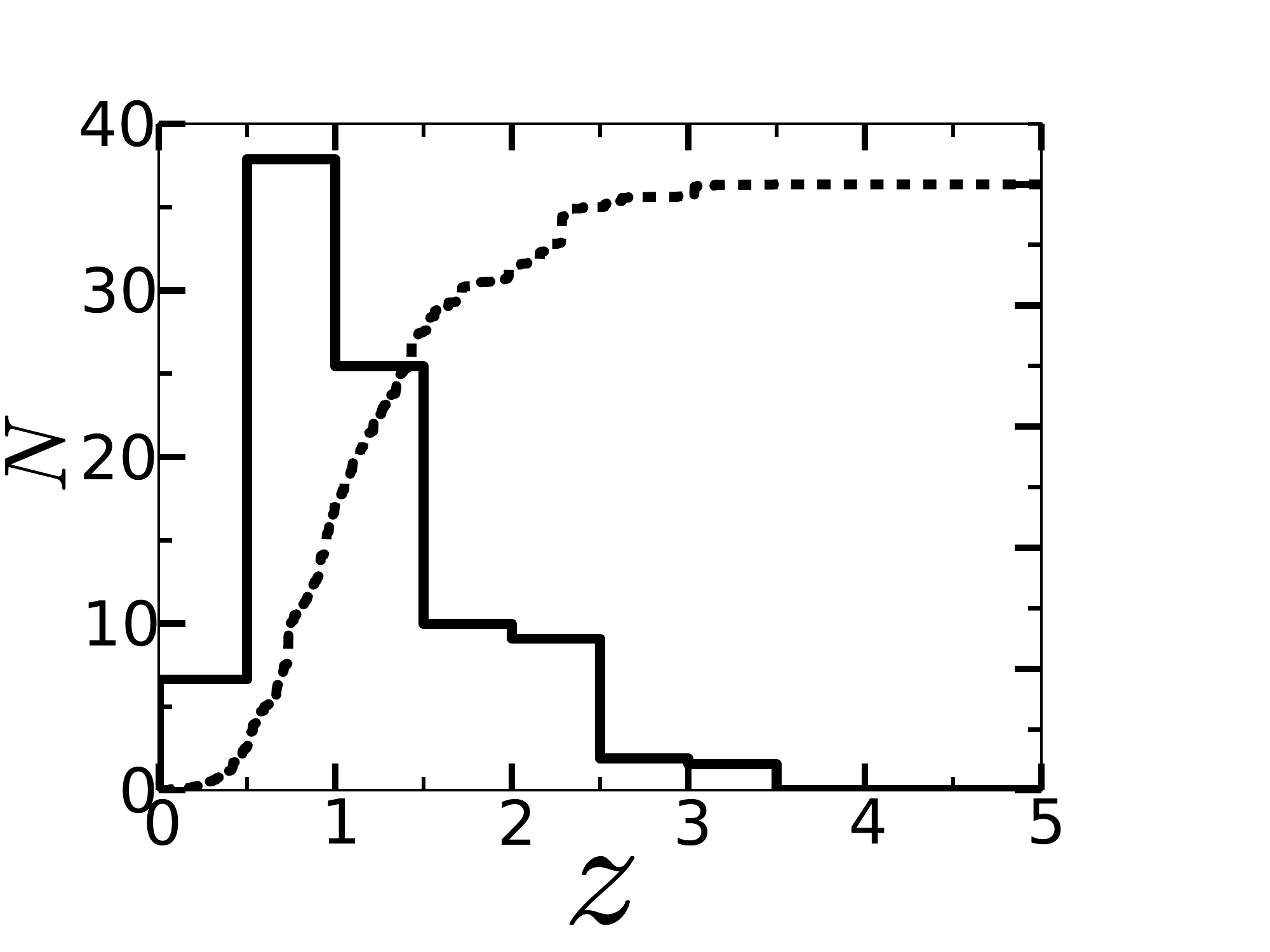}
\includegraphics[trim=0 10 55 0, clip, scale=0.22]{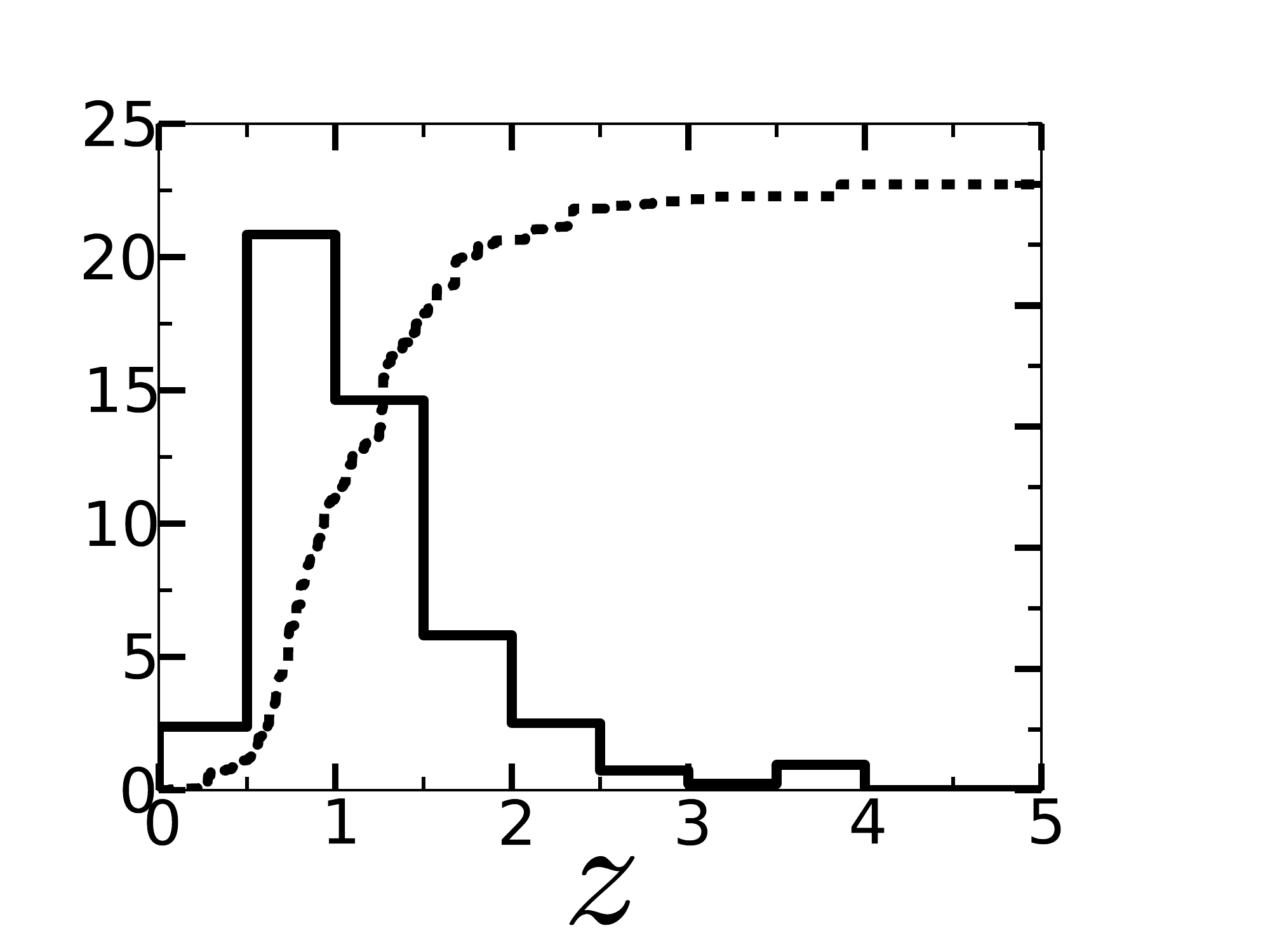}
\includegraphics[trim=0 10 55 0, clip, scale=0.22]{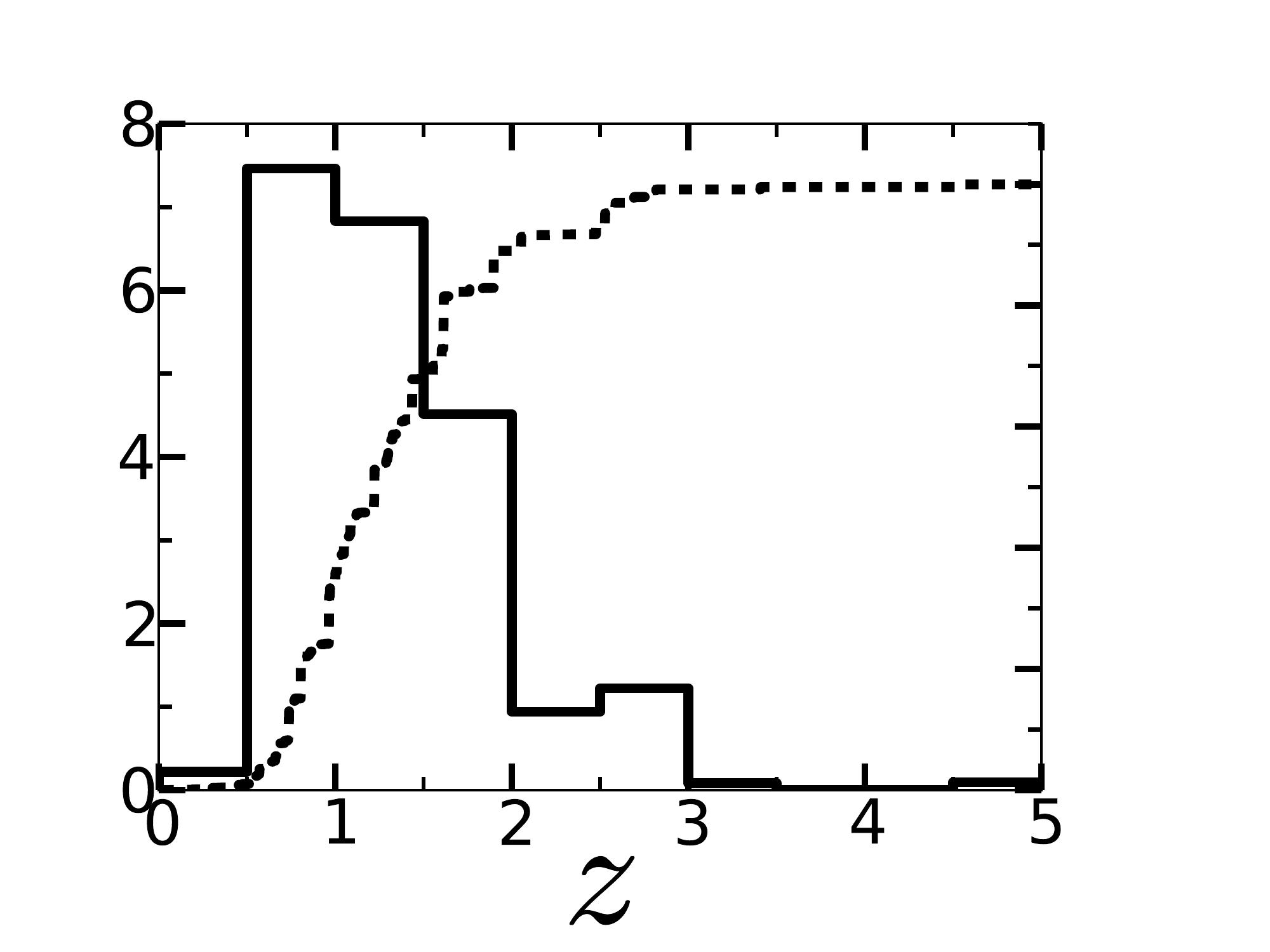}
\includegraphics[trim=0 10 10 0, clip, scale=0.22]{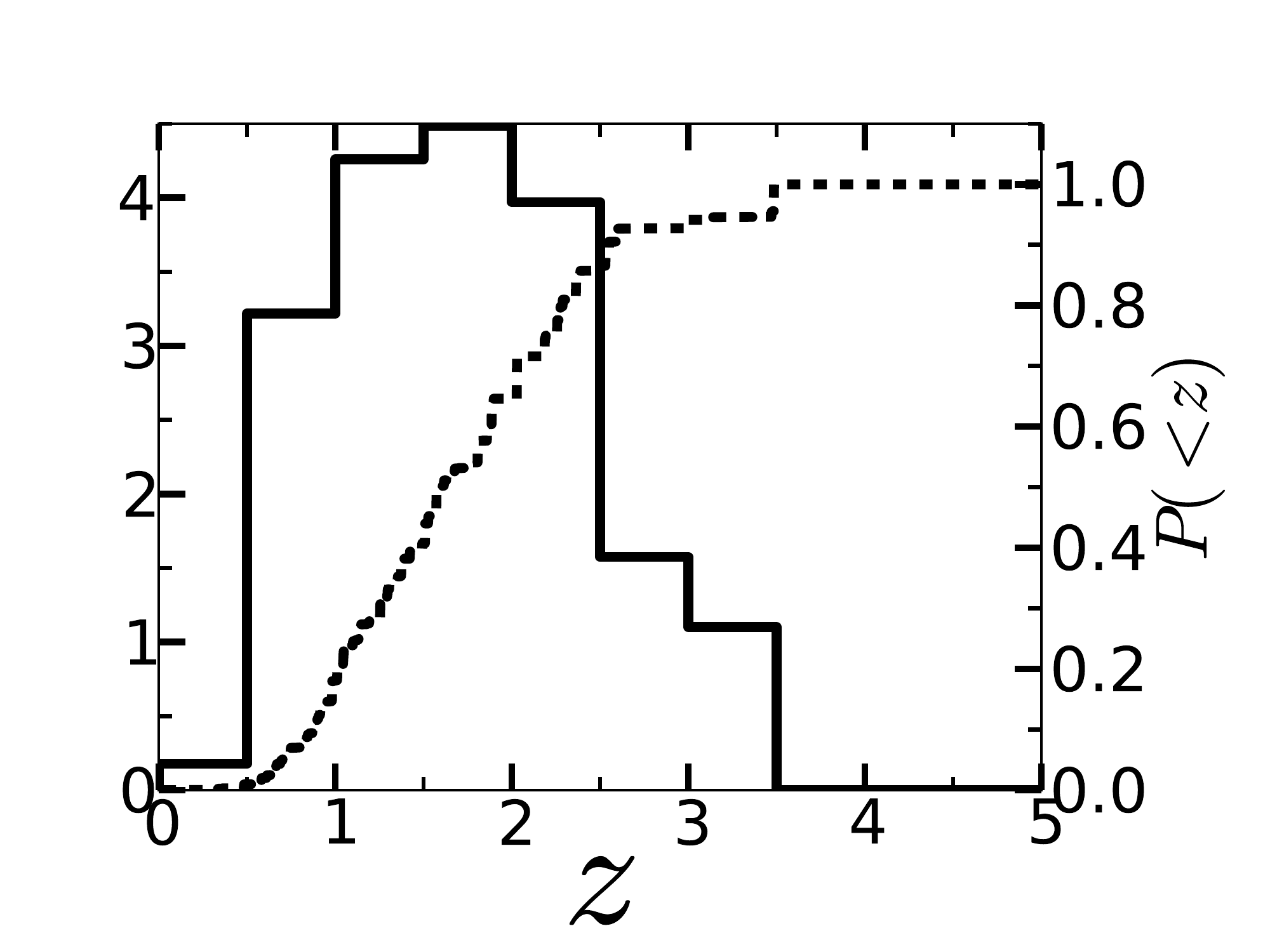}
\caption{The lensing-likelihood weighted distributions (solid) and cumulative distributions (dashed) of deflector properties. These distributions illustrate the diversity and evolution of the deflector population in the four samples analyzed. \textbf{Top row:} The distribution of $B$-band absolute magnitudes of the deflectors for the four LBG samples. \textbf{Middle row:} The distribution of image-deflector separations for the four LBG samples. \textbf{Bottom row:} The distribution of redshifts of the deflectors for the four LBG samples.}
\label{figure:lensprops}
\end{center}
\end{figure*}

\indent We present the distribution of the image-deflector separations, deflector redshifts and deflector $B$-band absolute magnitudes in this section. The number of lensed sources is weighted by the likelihood of lensing for each image-deflector configuration.\\
\indent The top row of Figure \ref{figure:lensprops} shows the distribution of lens rest-frame $B$-band magnitudes for each of the four independent LBG samples. The peak of the distribution occurs around $M_{B}\sim-22$ for each of the samples.\\
\indent The middle row of Figure \ref{figure:lensprops} shows the distribution of image-deflector separations for each of the LBG samples. The normalised cumulative fractions are shown as dashed lines. We observe an approximate increase in the peak of the separation distribution as redshift increases (from $\sim1\farcs0$ at $z\sim4$ to $\sim2\farcs0$ at $z\sim7$), consistent with the expectation that higher-redshift sources have larger deflection angles.\\
\indent The bottom row of Figure \ref{figure:lensprops} shows the distribution of deflector redshifts. The normalised cumulative fractions are shown as dashed lines. We observe an increase in the peak of the deflector redshift distribution from $z\sim4$ sources, where the deflector distribution peaks around $z\sim1$, to the $z\sim7$ sources where the peak occurs around $z\sim2$. This evolution is consistent with the expectation that lenses are most likely to be found at around half of the angular diameter distance to the source.\\

\section{Magnification Bias}
\label{section:magbias}
\begin{figure*}
\begin{center}
\includegraphics[trim=15 0 0 0, clip, scale=0.35]{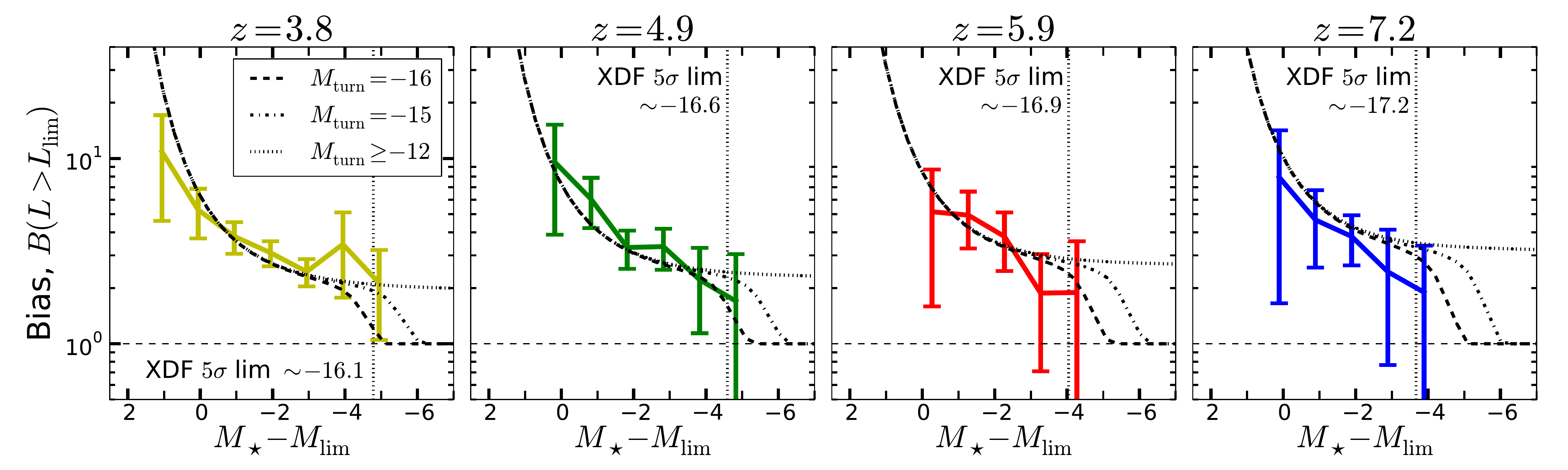}\\
\includegraphics[trim=15 0 0 30, clip, scale=0.35]{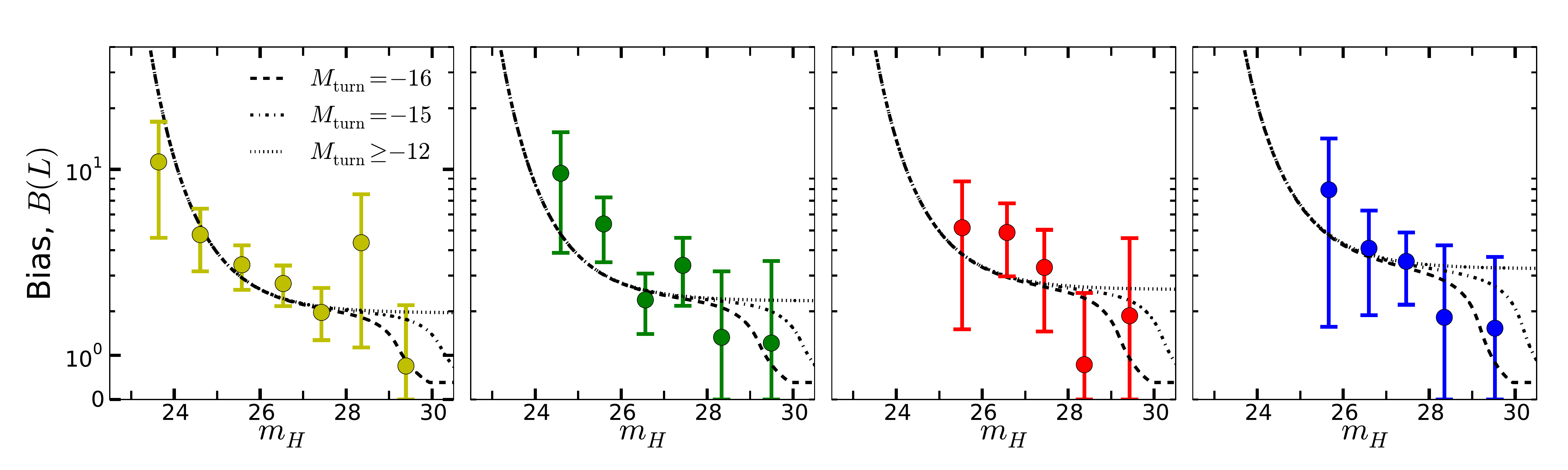}
\caption{\textbf{Top row:} The observed total magnification bias of all galaxies brighter than a flux limit $M_{\star}-M_{\mathrm{lim}}$ (solid) compared with the theoretical magnification bias for the luminosity functions (described by equation \ref{eqn:magbias2}) from \citet{bouwens2014uv} with a range of magnitudes at which the LF flattens ($\alpha\sim-1$), denoted by $M_{\mathrm{turn}}$. The cumulative lensed fraction is corrected for incompleteness according to the LF of \citet{bouwens2014uv}. \textbf{Bottom row:} The observed magnification bias in each bin plotted at the mean luminosity of the bin. At faint magnitudes in each sample of LBGs, the bias falls below the value expected from a faint-end slope continuing well-beyond the survey limit, indicating a possible deviation from a steep faint-end slope of the LF, although they agree with theory within their error bars. For a LF with a steep faint-end slope continuing well beyond the flux limit, the bias flattens to a value of $B\sim2-3$ (depending on $\alpha$). The measurements of bias at a fixed luminosities do not need to be corrected for incompleteness.}
\label{figure:magbias}
\end{center}
\end{figure*}

\indent The total magnification bias of a flux-limited sample, $B(L>L_{\textnormal{lim}})$, is the ratio of the fraction of strongly-lensed galaxies and the fraction of strongly-lensed random lines of sight, defined as the \textit{strong lensing optical depth}, $\tau$ \citep[e.g.][]{wyithe2011distortion}. We generate a catalogue of 50,000 random source positions in the GOODS fields and use the method of assessing the lensed fraction presented above in Section \ref{section:method} to determine the fraction of the source plane that will be strongly lensed. Based on our FJR, we assess the strong lensing optical depth for sources at $z\sim4$, $z\sim5$, $z\sim6$ and $z\sim7.2$ to be $\tau=0.41\%$, $0.54\%$, $0.65\%$, $0.75\%$. The values found are broadly consistent with theoretical predictions of the strong lensing optical depths at these redshifts \citep{mason2015correcting, wyithe2011distortion, barkana2000high}. Due to the large number of foregrounds in the CANDELS fields, the relative statistical uncertainty on $\tau$ is only $\sim4\%$ in all samples, and hence negligible in our bias calculations. We find consistent values for the optical depth if we apply a reasonable upper limit of $\sim350$kms$^{-1}$ on the inferred velocity dispersion of foreground galaxies. We find that our method of determining the optical depth returns values in close agreement with those in \citet{mason2015correcting} when we adopt their method of inferring velocity dispersions using stellar mass estimates\footnote{We use stellar mass estimates of foreground galaxies in the GOODS/CANDELS fields from the 3D-HST catalogue}. The optical depths are plotted as dashed lines in Figure \ref{figure:magfrac}.\\
\indent The bias is therefore the observed magnified fraction divided by the optical depth (the solid lines divided by the dashed lines in Figure \ref{figure:magfrac}). The observed total bias for each of the samples at a range of flux limits is plotted in the right panel of Figure \ref{figure:magfrac} and the top row of Figure \ref{figure:magbias}. The bias reaches values of $\sim10$ at bright magnitudes and high-redshifts, but near the survey flux limit has values lower than expected for a LF that remains steep well beyond survey limits.\\
\indent We calculate the observed magnification bias in each bin (as opposed to the total magnification bias for all galaxies brighter than a flux limit). The results are plotted in the bottom row of Figure \ref{figure:magbias}.\\
\indent For a LF with weak (or no) redshift-evolution of the $\alpha$ parameter, the magnification bias as a function of $M_{\star}-M_{\textrm{lim}}$ is expected to remain approximately constant with redshift. To highlight that this trend exists in the data, we plot the observed magnification bias at each redshift on the same axes in the right panel of Figure \ref{figure:magfrac}. While $\alpha$ evolves from $\sim-1.6$ to $\sim-2.0$ from $z\sim4$ to $z\sim8$, the statistical uncertainties in our measurements are larger than the change in bias from this evolution.\\
\indent For a given luminosity function, $\Psi(L)$, the magnification bias can be predicted analytically \citep{turner1984statistics} at luminosity $L$, by
\begin{equation}
B(L)=\frac{\int_{\mu_{\textrm{min}}}^{\mu_{\textrm{max}}}\frac{d\mu}{\mu}\frac{dP}{d\mu}\Psi(L/\mu)}{\Psi(L)},
\label{eqn:magbias1}
\end{equation}
where the $1/\mu$ factor accounts for the stretching of $d\mu$ with magnification, and $\frac{dP}{d\mu}$ is the magnification distribution for the brighter image in a strongly lensed system, given for an SIS by,
\begin{equation}
\frac{dP}{d\mu} = \left\{
\begin{array}{c l}
\frac{2}{(\mu-1)^{3}} & \quad \textrm{for $2<\mu<\infty$}\\
0 & \quad \textrm{for $\mu<2$}
\end{array}\right .
\label{eqn:magdist}
\end{equation}
We assume $\Psi(L)$ to be the Schechter luminosity function. The analytic magnification bias for all galaxies in a flux limited sample is
\begin{equation}
B(L>L_{\mathrm{lim}})=\frac{\int_{\mu_{\textrm{min}}}^{\mu_{\textrm{max}}}d\mu \int_{L_{\textrm{lim}}}^{\infty}dL\frac{dP}{d\mu}\Psi(L/\mu)}{\int_{L_{\textrm{lim}}}^{\infty}dL\Psi(L)}.
\label{eqn:magbias2}
\end{equation}
where the factor of $1/\mu$ is no longer included because we are now integrating over magnification, and the stretching of $d\mu$ no longer affects the integral.\\
\indent Results for our bias estimates given the Schechter LF parameters in \citet{bouwens2014uv} and theoretical curves are plotted in Figure \ref{figure:magbias}. The top row compares the theoretical bias of all galaxies in a flux limited sample with our measurements of the observed bias for all galaxies in a flux limited sample. The bottom row shows the theoretical bias of galaxies at a fixed luminosity with our measurements of the observed bias in each magnitude bin. Theoretical values for bias are calculated using previously derived LF parameters $\alpha$ and $M_{\star}$ \citep{bouwens2014uv} and a range of values at which the luminosity function deviates from a steep faint-end slope. We find close agreement between the observed shape and amplitude of the magnification bias and the theoretical function in each of the independent samples.\\
\indent It is worth noting that the inferred magnification bias is not sensitive to the parameters of the Faber-Jackson relation (Section \ref{section:FJR}), because the use of the FJR to determine the efficiency of observed galaxies affects both the numerator (fraction of strongly-lensed LBGs) and the denominator (the strong lensing optical depth) similarly.\\

\subsection{The Faint-end Slope Beyond Current Flux Limits}
\label{subsection:cutoff}
\begin{figure*}
\begin{center}
\includegraphics[trim=25 0 10 0, clip, scale=0.41]{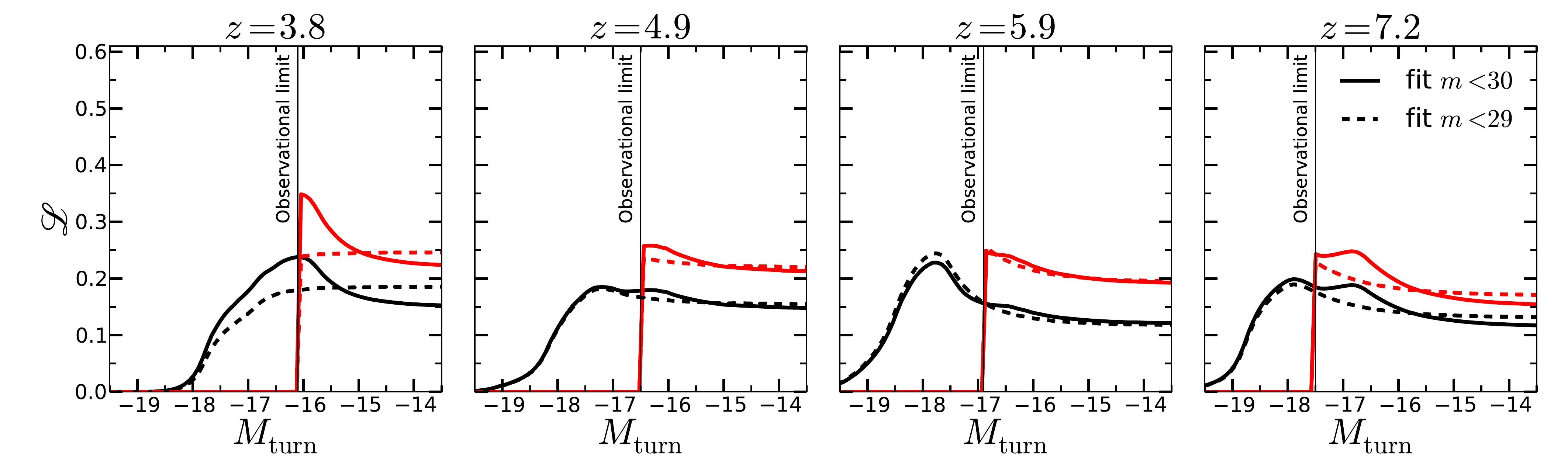}
\caption{The inferred value of $M_{\textrm{turn}}$ using the magnification bias measurements of all galaxies brighter than $m<30$ (solid) and all galaxies brighter than $m<29$ (dashed). The bias measurements which we fit to are shown in the bottom row of Figure \ref{figure:magbias}. In black we plot the likelihoods with a prior on $\alpha$ and $M_{\star}$ from \citet{bouwens2014uv} (see Table 4 therein for values). The red curves show the estimated likelihoods including an additional requirement that the minimum magnitude is fainter than the magnitude to which current observations confirm a steep faint-end slope. We find approximately consistent preferred values of $M_{\textrm{turn}}$ in each of the four samples, but with varying amplitudes. For LBGs brighter than $m=30$ (approximately the $5\sigma$ XDF limit, solid line), we find a preferred value of $M_{\textrm{turn}}$ around the observational limits in each sample. However, a value of $M_{\mathrm{turn}}>-16$ is not excluded. For only LBGs brighter than $m=29$ (dashed line), we find no constraint on $M_{\textrm{turn}}$ in any of the samples. This is expected, because to constrain $M_{\textrm{turn}}$ we need to consider galaxies within $\sim1-2$ mag of $M_{\textrm{turn}}$. The curves are normalised such that the probability of $M_{\textrm{turn}}<-12$ is unity.}
\label{figure:lmin}
\end{center}
\end{figure*}

\begin{figure}
\begin{center}
\includegraphics[trim=0 0 0 0, clip, scale=0.33]{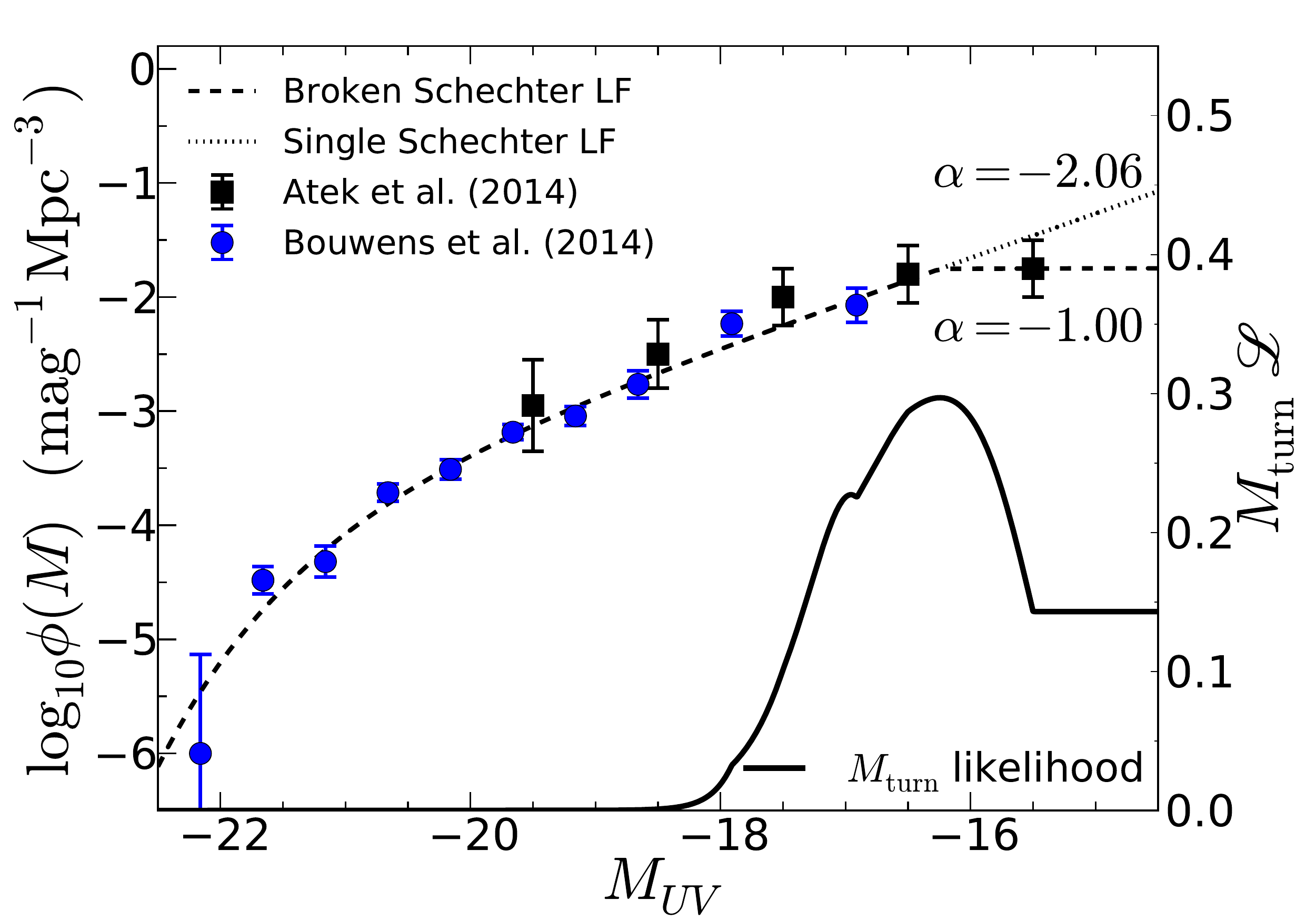}
\caption{The $z\sim7$ LF data from \citet{bouwens2014uv} (blue circles) and \citet{atek2014new} (black squares) with the best fit broken Schechter LF (dashed), and single Schechter LF (dotted). The solid line shows the likelihood of $M_{\mathrm{turn}}$ given the two datasets. We find a favoured flattening magnitude at $M_{UV}\sim-16.5$, consistent with our magnification bias measurements.}
\label{figure:atekLF}
\end{center}
\end{figure}

\indent Magnification bias results from magnification of intrinsically faint sources below an observed flux limit into an observed sample, hence quantifying the degree of magnification bias offers an opportunity to investigate the behaviour of the LF beyond current survey limits.\\
\indent To illustrate, we begin with a toy model in which there is a minimum luminosity for galaxies of $L_{\textrm{min}}$, below which there are no galaxies, and a power-law slope of $\alpha=-2.0$ for $L>L_{\textrm{min}}$. In this toy model, a sharp cutoff in the LF at a value of $L_{\textnormal{min}}$ yields a bias of,
\begin{equation}
B(L>L_{\textnormal{lim}}) = \left\{
\begin{array}{l l}
3 - \frac{2}{L_{\textnormal{lim}}/L_{\textnormal{min}} - 1} & \quad \textrm{for $L_{\textnormal{lim}}>2L_{\textnormal{min}}$}\\
1 & \quad \textrm{for $L_{\textnormal{min}}<L_{\textnormal{lim}}<2L_{\textnormal{min}}$}
\end{array}\right .
\end{equation}
This implies the total bias of a flux-limited sample reaches unity approximately $1$ mag brighter than $L_{\textrm{min}}$. We observe a hint of the possibility of this occurring in the four samples presented in this paper, as seen in the top row of Figure \ref{figure:magbias}. The bottom row of Figure \ref{figure:magbias} also highlights this behaviour in our samples.\\
\indent Rather than a sharp cutoff in the LF, we consider a more realistic model with a LF that flattens ($\alpha_{2}=-1$) after some luminosity, $L_{\mathrm{turn}}$. We attempt to constrain $L_{\textrm{turn}}$ by finding a LF that will reproduce the observed magnification bias in the bottom row of Figure \ref{figure:magbias}. Using a broken Schechter function of the form,
\begin{equation}
\Psi(L)dL= \left\{
\begin{array}{l l}
\Psi_{\star,1}\Big(\frac{L}{L_{\star}}\Big)^{\alpha_{1}}\textrm{exp}\Big(-\frac{L}{L_{\star}}\Big)\frac{dL}{L_{\star}} & \quad \textrm{for $L\geq L_{\textrm{turn}}$}\\
\Psi_{\star,2}\Big(\frac{L}{L_{\star}}\Big)^{\alpha_{2}}\textrm{exp}\Big(-\frac{L}{L_{\star}}\Big)\frac{dL}{L_{\star}} & \quad \textrm{for $L<L_{\textrm{turn}}$}
\end{array}\right.
\end{equation}
with $\alpha_{2}=-1$ (i.e. a flat LF beyond $L_{\textrm{turn}}$), we fit the bias calculated from equation \ref{eqn:magbias1} to the data with $L_{\textrm{turn}}$, $\alpha_1$, and $M_{\star}$ as free parameters. We include priors on the values of $\alpha_1$ and $M_{\star}$, from the LFs of \citet{bouwens2014uv}. Figure \ref{figure:lmin} shows the constrains found for the minimum luminosity from our analysis fitting to two subsets of our measurements. The dashed black line in Figure \ref{figure:lmin} shows the probability distribution function (PDF) when fitting $M_{\textrm{turn}}$ to the observed bias of only LBGs brighter than $m=29$, and the solid line shows the PDF when fitting to LBGs brighter than $m=30$ (approximately the $5\sigma$ limit in the XDF). The PDFs are normalised such that the probability of $M_{\textrm{turn}}<-12$ is unity. We find preferred values of $M_{\mathrm{turn}}$ to peak around the current observational limits in each of the independent samples from $4<z<7$. The sample at $z\sim6$ peaks at a magnitude brighter than flux limits, which can be ruled out observationally. We also calculate the likelihoods with an additional prior enforcing $M_{\mathrm{turn}}$ to occur below the magnitude that the steep faint-end slope has been observed to extend to. These are plotted in red in Figure \ref{figure:lmin} for the same flux limits as above. It is important to note that the inference of $M_{\mathrm{turn}}$ occurring close to current flux limits is marginal and does not rule out a faint-end slope extending well-beyond current flux limits, or a flattening for a few magnitudes followed by an upturn.\\
\indent We find that the constraint disappears when $M_{\mathrm{turn}}$ is fit to only brighter ($m<29$) galaxies. We extend this test by recalculating our results by omitting XDF LBGs entirely from the analysis to investigate whether the observed magnification bias will always approach unity near the flux limit due to selection effects. When we perform this test, we find the magnification bias of galaxies in the GOODS-North and GOODS-South is completely consistent with that of the full sample, rather than approaching unity near the flux limit.\\
\indent It is important to note that the magnification bias is only observed to drop below its expected value close to the current flux limits where selection effects become significant. While we have taken care to account for the decreased sensitivity to very faint sources, there still exists the possibility that we have missed a significant fraction of gravitationally lensed LBGs at very faint magnitudes. Furthermore, if the interloper fraction at very faint fluxes is high, the lensed fraction will be underestimated, causing a spurious inference of $M_{\mathrm{turn}}$.\\
\indent The possibility of a flattening of the LF at $M_{\mathrm{turn}}\sim-16.5$ at $z\sim7$ is consistent with observations of LBGs down to $M_{UV}\sim-15.5$ of magnified $z\sim7$ LBGs using Frontier Fields cluster Abell 2744 \citep{atek2014new, ishigaki2014hubble}. Figure \ref{figure:atekLF} shows the \citet{bouwens2014uv} and \citet{atek2014new} data with the best fit broken Schecter function ($\alpha_2=-1$). We find that a broken Schechter function represents the data very well, and offers an independent constraint on $M_{\mathrm{turn}}$. The data favours a value of $M_{\mathrm{turn}}\sim-16.5$, consistent with $M_{\mathrm{turn}}$ inferred from our magnification bias results. However, as is the case with our magnification bias analysis, this inference is based on a single data point.\\
\indent Further to this study at $z\sim7$, there exist measurements of the UV luminosity function of $z\sim2$ LBGs down to $M_{UV}\sim-13$ \citep{alavi2014ultra}. While a single Schechter function favours a steep slope extending to $M_{UV}\sim-13$ at $z\sim2$, the shape of the LF at the faint end may be more complicated than this simple parametrisation. In fact, the data also seems to suggest a flattening of the population density between $-17\lesssim M_{UV} \lesssim-15$, before a steeper upturn from $-15\lesssim M_{UV} \lesssim-13$ \citep[see][Figure 7]{alavi2014ultra}. This more complicated LF would produce a reduced magnification bias ($B(L>L_{\mathrm{lim}})\sim1$) of $z\sim2$ LBGs near a flux limit of $M_{UV}\sim-17.5$, which is on the edge of current survey limits in blank fields \citep{sawicki2012stars, oesch2010evolution, hathi2010uv, reddy2009steep}. Additionally, a flattening, and potentially a rise in the LF at $M_{UV}\gtrsim-15$ would not be inconsistent with the inference from GRB host galaxies studies \citep{trenti2012gamma,trenti2013gamma,tanvir2012star}. In fact, these studies only constrain the presence of an abundant population of galaxies below the XDF detection limit, but not the shape of the galaxy LF, which they \emph{assume} to be Schechter-like. Interestingly, theoretical models that are based on a double population of faint galaxies have been proposed in the context of hydrogen reionization (e.g., see \citealt{alvarez2012constraints}). \\

\subsection{Deriving Schechter Parameters from Lensing}
\label{subsection:lfparams}
\begin{figure*}
\begin{center}
\includegraphics[trim=25 0 10 0, clip, scale=0.42]{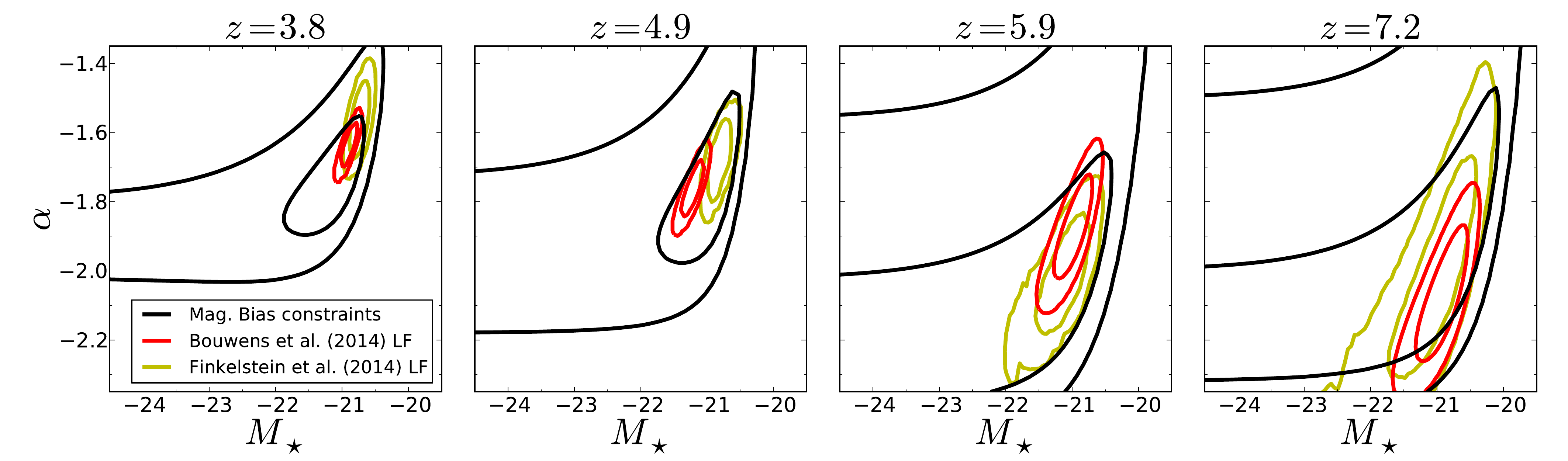}
\caption{Measurement of the Schechter parameters, $\alpha$ and $M_{\star}$ using only the observed magnification bias as a function of $M_{\star}-M_{\mathrm{lim}}$ (black). The contours from \citet{bouwens2014uv} (red) and \citet{finkelstein2014evolution} (yellow) are shown for comparison. We find close agreement between the two methods at $z\sim4$ and $z\sim5$, while the constraints at $z\sim6$ and $z\sim7$ from lensing are weaker due to the larger error bars on the magnification bias measurements.}
\label{figure:lfparams}
\end{center}
\end{figure*}

\indent A very interesting application of this analysis is that Schechter function parameters $\alpha$ and $M_{\star}$ can be derived directly from the magnification bias. This method is completely independent of the standard procedure using number counts of galaxies, and therefore could be combined to produce improved constraints.\\
\indent The magnification bias at a fixed luminosity can be predicted using equation \ref{eqn:magbias1}, and is a function of $\alpha$ and $M_{\star}$. By fitting the predicted bias of LBGs at a fixed flux to our measured bias in each flux bin (which is shown in the bottom row of Figure \ref{figure:magbias}, and is \textbf{not} the LF-corrected cumulative fraction), we can constrain $\alpha$ and $M_{\star}$\footnote{We fit the theoretical bias at the mean magnitude of the LBGs in each magnitude bin to the observed bias in that bin with $\alpha$ and $M_{\star}$ as free parameters}. This does not rely on any prior knowledge of the LF. Because $\alpha$ and $M_{\star}$ are much more sensitive to the bias of bright galaxies than that of faint galaxies, and we see a possible deviation from a single Schechter function at faint magnitudes, we exclude the two faintest bins ($29<m<30$, and $28<m<29$) from the fit\footnote{In fact, fitting the data including the two faintest bins with an extra free parameter, $M_{\mathrm{turn}}$, and marginalising over this parameter gives the same result}. Figure \ref{figure:lfparams} shows the constraints on the LF from the observed magnification bias alone along with the constraints from number counts of the same samples.\\
\indent The measurements of the bias at $z\sim4$ and $z\sim5$ do an excellent job of constraining the luminosity function. At higher redshift, as the samples become smaller and the random errors grow, we cannot constrain the LF as effectively. However we find that our observations are consistent with the UV luminosity functions presented by \citet{bouwens2014uv}. This also provides an internal consistency check of our analysis.\\

\subsection{Contaminant Discussion}
\label{subsection:contaminants}
\indent We check for bias arising from the selection of LBGs. There may be an enhancement and reddening of LBG candidates observed around bright, red foreground galaxies due to photometric scatter, causing an increased fraction of interlopers around such foreground objects and providing a false lensing signal among bright candidates. To determine if this may affect our results, we check if an enhanced interloper fraction around bright foregrounds is found in lower-redshift LBG samples for which there exists spectroscopic follow up. We combine catalogues with spectroscopically confirmed LBGs and photometrically selected LBGs which were identified as interlopers from $3<z<6$ using observations reported by \citet{vanzella2009spectroscopic, reddy2006spectroscopic, malhotra2005overdensity, steidel2003lyman}. We compare the fraction of interlopers for LBGs within $5\farcs0$ of bright, red foreground galaxies ($m_{r}<-22$ mag) with the fraction of interlopers in the total sample. In both cases, we find the interloper fraction to be $\sim8\%$, with $2$ of $25$ LBGs around bright foreground galaxies identified as interlopers, and $21$ of $252$ of the entire sample. This indicates that the alignment between bright LBGs and massive foreground galaxies is not likely due to selection bias. The large enhancement in false identifications required to mimic the observed magnification bias of $\sim10$ for bright galaxies is clearly inconsistent with the spectroscopic data.\\

\section{Magnification Bias and the Luminosity Function}
\label{mag-lumfunc}
\begin{figure}
\includegraphics[scale=0.35]{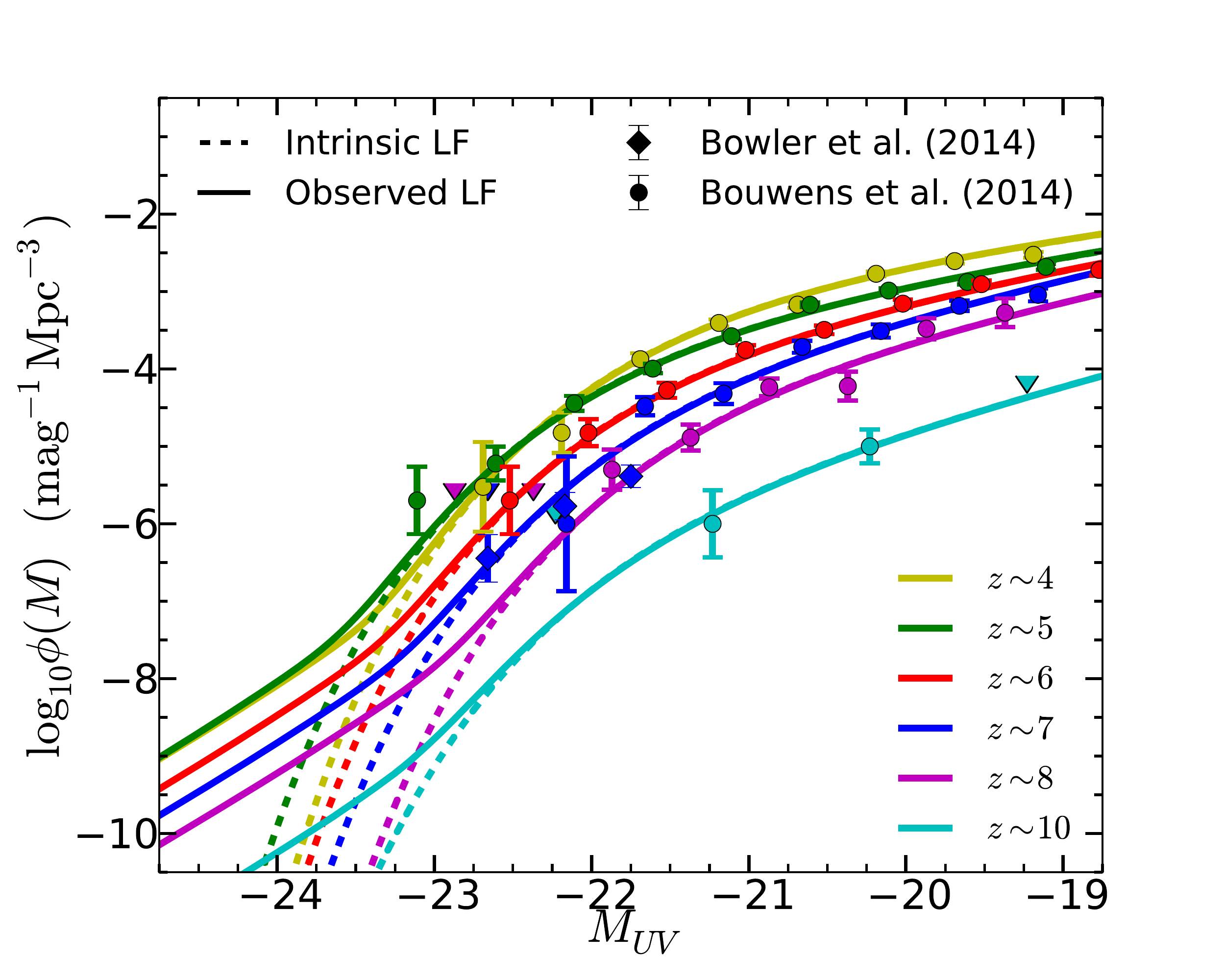}
\caption{The effect of magnification bias on the bright end of the luminosity functions $z\sim4, z\sim5, z\sim6, z\sim7, z\sim8$ and $z\sim10$ (LBGs become monotonically more abundant with decreasing redshift at $M_{UV}\leq-22$). The observed luminosity functions are shown as solid lines, and the intrinsic luminosity functions as dashed lines. The luminosity function measurements from \citet{bouwens2014uv} and \citet{bowler2014bright} are plotted as circles and diamonds, respectively. At $z\sim7, z\sim8$ and $z\sim10$, observations are close to probing the bright end where gravitational lensing becomes a significant effect, but not bright enough for it to be manifested in the observed LF.}
\label{figure:obsLF}
\end{figure}

\indent The effect of magnification bias on determining the luminosity function is an important consideration when making a census of galaxies in the epoch of reionization \citep{mason2015correcting, wyithe2011distortion}. In this section we show the effect that the lensed fraction reported in this paper has on the observed luminosity function.\\
\indent The observed luminosity function of LBGs is the convolution between the intrinsic luminosity function, $\Psi(L)$, and the magnification distribution of an SIS, $\frac{dP}{d\mu}$, weighted by the strong lensing optical depth, $\tau$. This results in an observed LF, $\Psi_{\textrm{obs}}(L)$, with a power law tail at the bright end with a slope of $-3$ (the slope of the magnification distribution of an SIS). We assess the affect of gravitational lensing on the LF by following the method presented by \citet{wyithe2011distortion}, where it is modelled by considering the optical depth, $\tau$, the mean magnification of multiply-imaged sources for an SIS, $\langle\mu_{\textrm{mult}}\rangle=4$, and the demagnification of unlensed sources (to conserve total flux on the cosmic sphere), $\mu_{\textrm{demag}}=(1-\langle\mu_{\textrm{mult}}\rangle\tau)/(1-\tau)$. The observed LF is then given by
\begin{equation}
\begin{split}
\Psi_{\textrm{obs}}(L)=& (1-\tau)\frac{1}{\mu_{\textrm{demag}}}\Psi(L/\mu_{\textrm{demag}})\\
&+\tau\int_{0}^{\infty}d\mu\frac{1}{\mu}\Big(\frac{dP_{\textrm{m,1}}}{d\mu}+\frac{dP_{\textrm{m,2}}}{d\mu}\Big)\Psi(L/ \mu),
\end{split}
\end{equation} 
where $\frac{dP_{\textrm{m,2}}}{d\mu}=2/(\mu+1)^{3}$ for $0<\mu<\infty$ is the second image's magnification probability distribution, and $\frac{dP_{\textrm{m,1}}}{d\mu}$ is given by equation \ref{eqn:magdist}. We use the values of the optical depth from our analysis presented in Section \ref{section:magbias}, and calculate the optical depth at $z=6.8$, $z=7.9$ and $z=10.4$ to be $\tau=0.72\%$, $0.80\%$ and $0.94\%$, respectively.\\
\indent We begin by assuming that the observed luminosity function is not affected by gravitational lensing and hence represent the intrinsic luminosity functions. We plot these intrinsic luminosity functions \citep{bouwens2014uv}, the inferred observed luminosity function, and observations \citep{bowler2014bright, bouwens2014uv} at $z\sim4, z\sim5, z\sim6$, $z\sim7$, $z\sim8$ and $z\sim10$ in Figure \ref{figure:obsLF} (dashed lines). Figure \ref{figure:obsLF} also shows the biased LFs, illustrating the luminosity at which gravitational lensing becomes important. This also illustrates the assumption that current LF measurements are not significantly affected by magnification bias is sound.\\
\indent The effect of magnification bias is not significant at the faint end of the luminosity function. At around $2$ magnitudes fainter than $M_{\star}$, the excess observed abundance of LBGs is on the order of $0.5\%$ for all of the samples, which is significantly smaller than the observational errors in the abundances at these magnitudes.\\
\indent We note that even the brightest \citet{bowler2014bright} and \citet{bouwens2014uv} measurements are not bright enough to probe the affected region of the LF. However, the effect magnification bias will have on surveys at $z\gtrsim8$ is obvious from the solid lines for $z\sim8$ and $z\sim10$ where the observed LF will display a break from the intrinsic LF around $M_{UV}\lesssim-22.5$.\\
\indent Not plotted in Figure \ref{figure:obsLF} are extrapolated LFs at $z>10$. \citet{wyithe2011distortion} showed that if $M_{\star}$ drops sharply at high redshifts, surveys of the depth of the XDF with JWST will observe galaxies at $z>10$ in the affected region of the LF.\\

\section{Analysis of current $J_{125}$-dropouts}
\label{Jdrops}
\indent In Section \ref{mag-lumfunc}, we presented the effect that magnification bias has on the observed luminosity function. Figure \ref{figure:obsLF} highlights that while magnification bias is not a significant effect in current surveys out to $z\sim8$, the affected region of the $z\sim10$ LF begins at around $M_{UV}\sim-22.5$. We investigated the $4$ unusually bright $z\sim10$ $J_{125}$-dropouts presented by \citet{oesch2014most} to search for evidence of lensing. This point is discussed in \citet{oesch2014most}, where they find there is the possibility of a modest amount of lensing in $2$ of the $4$ dropouts. By applying the technique employed in this paper, we assign likelihoods of lensing to the four $z\sim9-10$ LBGs.\\
\indent As noted by \citet{oesch2014most}, $2$ of the LBGs are not close in projection to any foreground objects \citep[\textit{GN-z10-3} and \textit{GN-z9-1} in the notation of][]{oesch2014most}, while the other two do have projected neighbours (\textit{GN-z10-1} and \textit{GN-z10-2}). \textit{GN-z10-1} is $1\farcs2$ from a foreground galaxy at $z_{phot}=1.6$ with $M_{B}=-20.1$ (using photometry from the 3D-HST catalogue). \citet{oesch2014most} infer a photometric redshift of $z=1.8$). Using our redshift-evolving FJR, this corresponds to a stellar velocity dispersion of $140$ kms$^{-1}$. The required stellar velocity dispersion for strong-lensing in this case is $198$ kms$^{-1}$, giving this LBG a likelihood of lensing of $\mathscr{L}=0.12$. \textit{GN-z10-2} is $2\farcs9$ from a bright galaxy at $z_{spec}=1.02$ with $M_{B}=-20.7$. This corresponds to an inferred stellar velocity dispersion of $214$ kms$^{-1}$. The required stellar velocity dispersion for strong lensing is $279$ kms$^{-1}$, giving a likelihood of lensing of $\mathscr{L}=0.1$. While the statistics are too small to draw any firm conclusions, this average observed lensed fraction of $\sim6\%$ for the four galaxies is consistent with a lensed fraction of LBGs brighter than $M_{\star}$ of $\sim10\%$.\\

\section{Summary}
\label{section:conclusion}
\indent We have estimated the likelihood of strong gravitational lensing of LBGs in the XDF and GOODS at $z\sim4$, $z\sim5$, $z\sim6$ and $z\sim7$. We used a calibrated Faber-Jackson relation to estimate the lensing potential of all foreground objects in the fields. The result is a measurement of significant magnification bias in current high-redshift samples of LBGs. Our analysis allows us to draw the following conclusions,
\begin{enumerate}
\item Approximately $6\%$ of LBGs at $z\sim7$ brighter than $M_{\star}$ ($m_{H_{160}}\sim26$ mag) are expected to have been strongly gravitationally lensed with $\mu>2$;\\
\item The observed strongly lensed fraction of LBGs at all values of $m_H$ falls monotonically from $z\sim7$ to $z\sim4$, which can be explained by the expected evolution in the optical depth with redshift, and also $M_{\star}$ appearing brighter at lower redshift;\\
\item By evaluating the optical depth in our lensing framework, we calculate the magnification bias in each sample as a function of $M_{\star}-M_{\textrm{lim}}$, and find that the results agree at each redshift and are well described by theoretical predictions;\\
\item Extrapolation of our analysis leads to expectations for an increased fraction of strongly lensed galaxies at $z\gtrsim 8$, consistent with \citet{wyithe2011distortion};\\
\item The magnification bias of the faintest LBGs in the sample suggests there may be a flattening of the faint-end slope below current detections limits ($M_{UV}\gtrsim -16.5$). However, this result relies on LBG detections at low S/N in the XDF, and the constraints are weak. We present this result tentatively, with deeper data needed to better understand the population of faint high-$z$ galaxies; \\
\item Assessing the magnification bias as a function of luminosity offers an independent method of determining Schechter parameters $\alpha$ and $M_{\star}$. The results from this method are consistent with those found by fitting the LF based on number counts.
\end{enumerate}
%
%
\indent With the confirmation of the role of magnification bias come important consequences for future surveys of galaxies at $10< z< 20$. Currently, the LFs at $z\sim8$ need not be corrected for magnification bias.  However, magnification bias will be significant for luminosity functions at $z\gtrsim10$, notably in the JWST era \citep{wyithe2011distortion}.  In particular, with $M_{\star}$ possibly dropping rapidly beyond $z\sim8$ \citep{oesch2014most}, JWST will identify predominantly gravitationally lensed galaxies at $z\gtrsim10$.\\

{\bf Acknowledgments} 
JSBW is supported by an Australian Research Council Australian Laureate Fellowship.
MT acknowledges support from the Australian Research Council through the award of a Future Fellowship.
TT acknowledges support from the Packard Foundation via a Packard Fellowship.
We thank Charlotte Mason for her useful comments and conversations.
We thank Adriano Fontana for his useful comments and suggestions.

\bibliographystyle{mn2e}
\bibliography{magbias}

\appendix
\section{Spatial Correlations Between Bright Foregrounds \& LBGs}
\label{appendix:spatial}
\begin{figure*}
\includegraphics[trim=0 0 25 0, clip, scale=0.245]{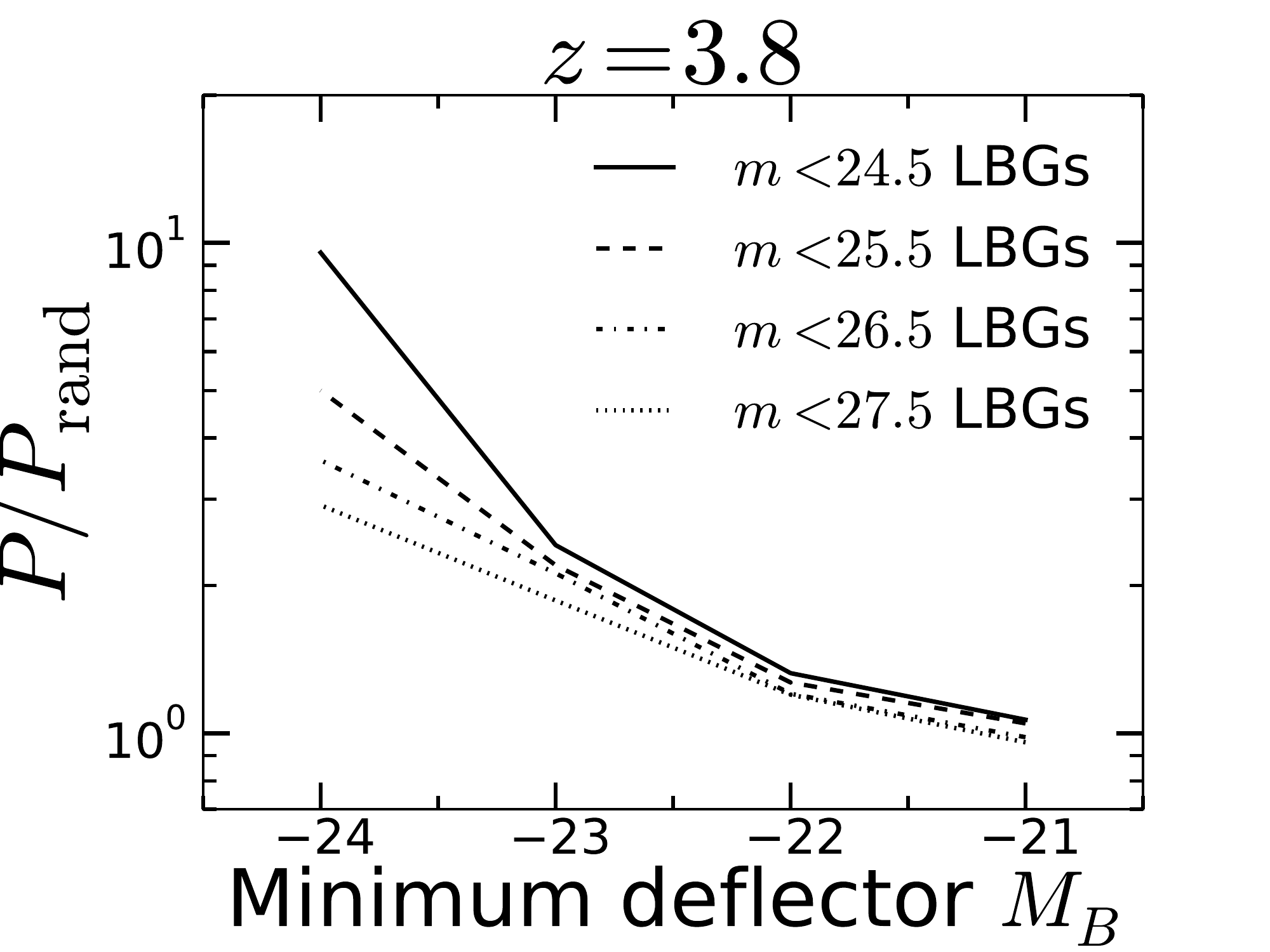}
\includegraphics[trim=83 0 25 0, clip, scale=0.245]{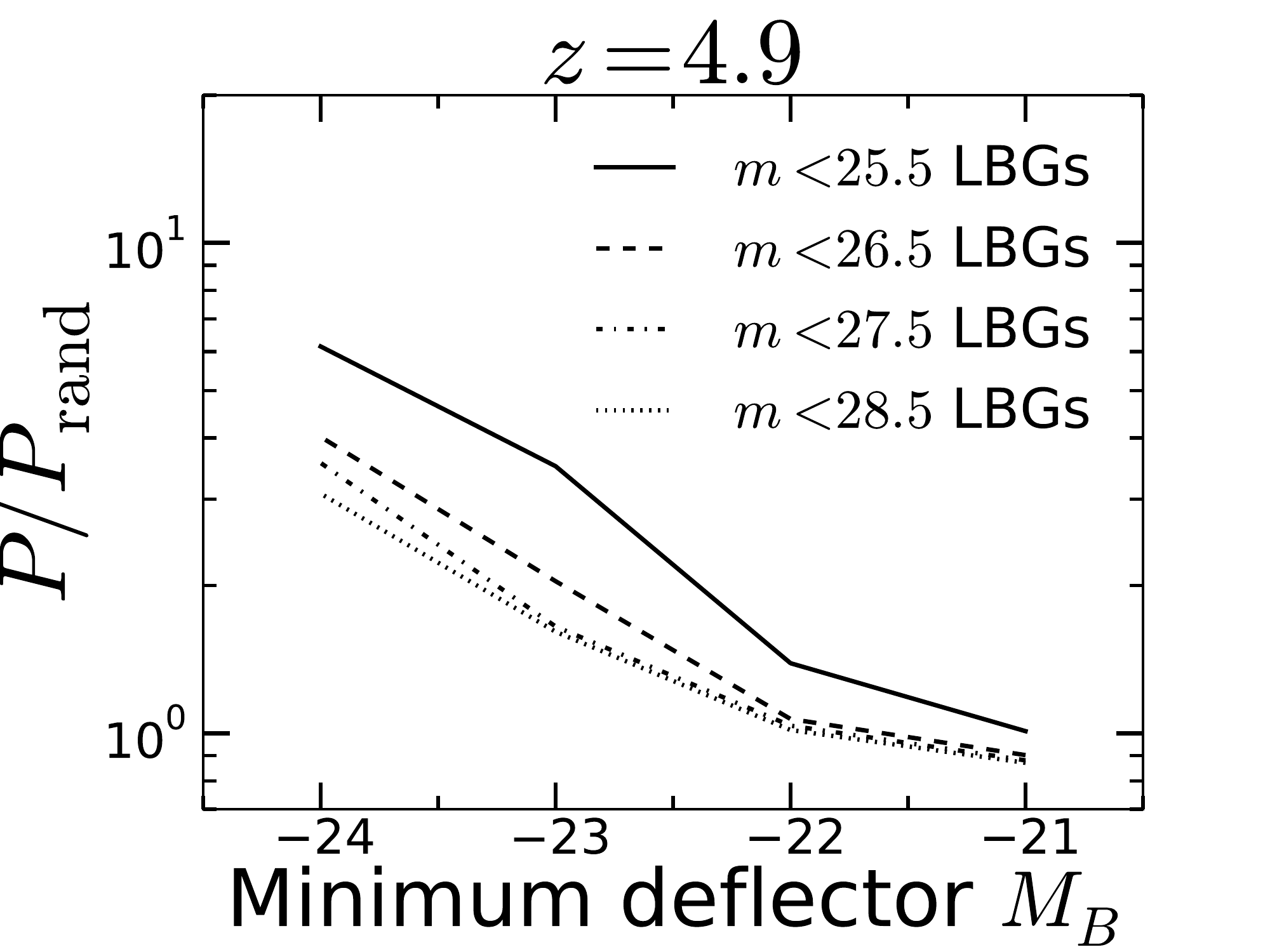}
\includegraphics[trim=83 0 25 0, clip, scale=0.245]{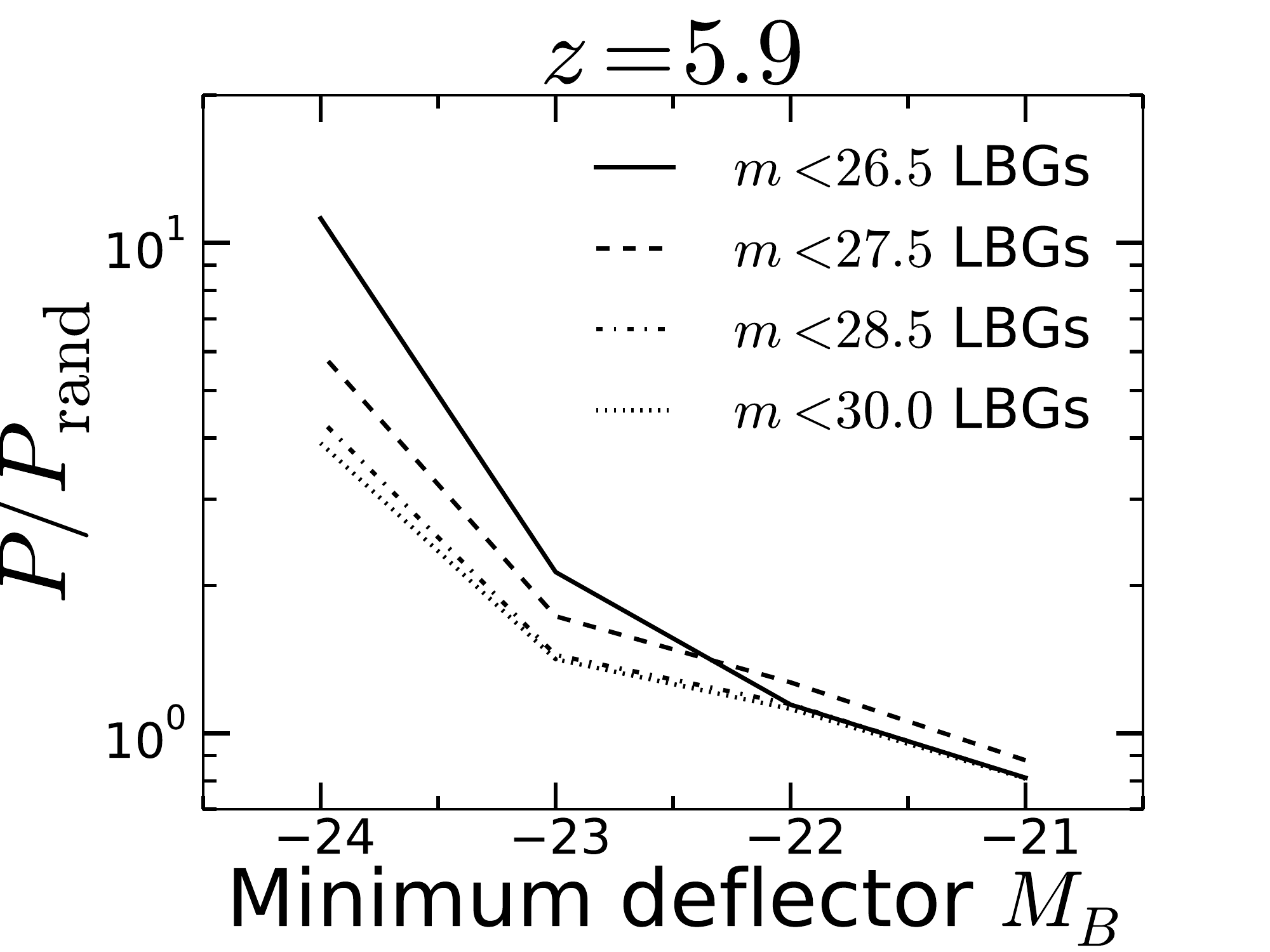}
\includegraphics[trim=83 0 25 0, clip, scale=0.245]{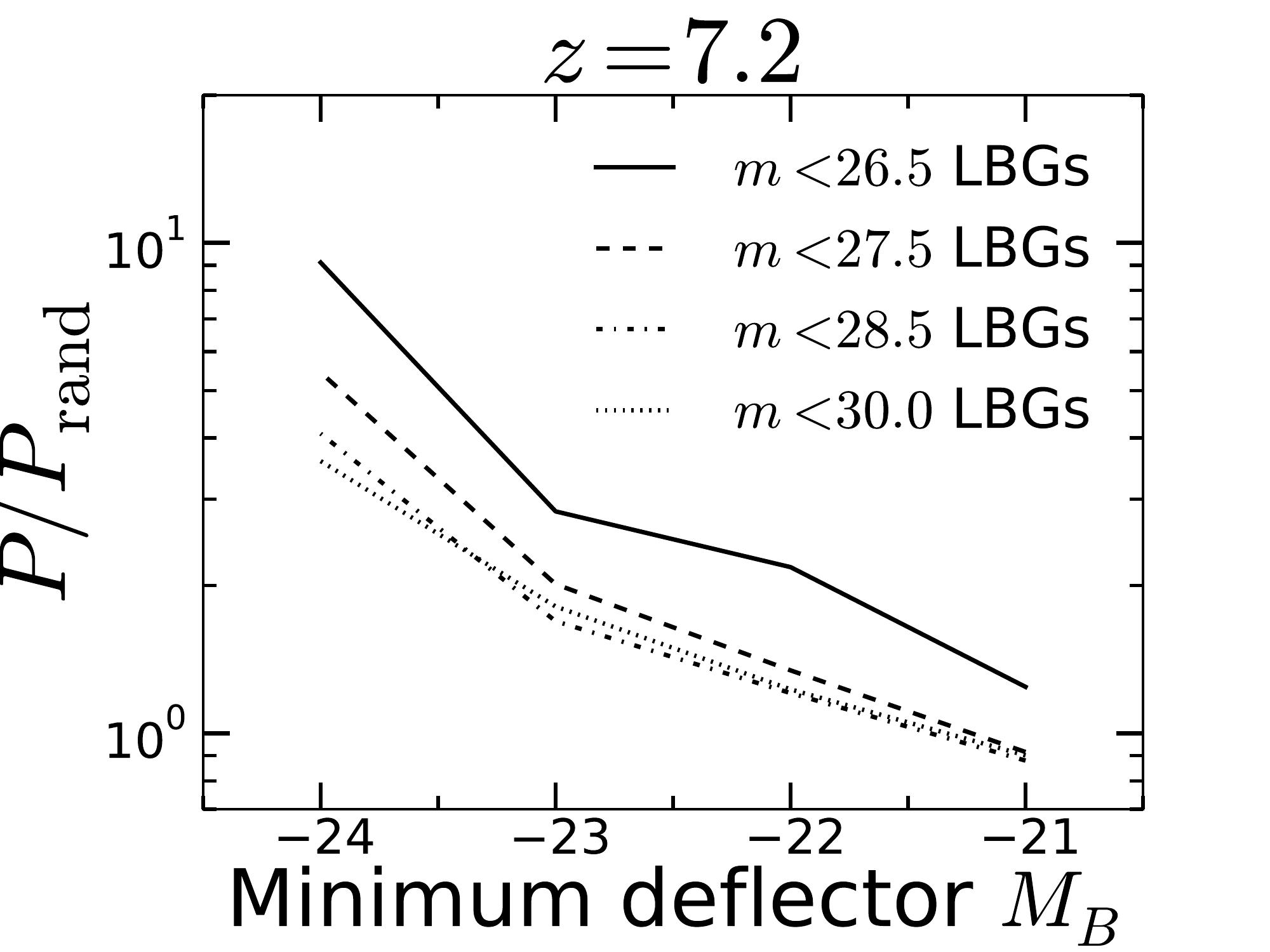}
\caption{The excess probability of finding an LBG brighter than various flux limits within $5\farcs0$ of deflectors brighter than $M_{B}$ at $z<2$. In each of the four samples, we find that there is an excess of LBGs around bright foreground objects. The excess becomes monotonically more pronounced with brighter flux limits around bright foregrounds in each of the four samples. At each redshift slice, we consider a different set of flux limits as $M_{\star}$ appears brighter for the lower-redshift samples. The right panel shows a large excess of bright LBGs at $z\sim7$ around bright foreground objects. At $z\sim4$, we find similar behaviour of bright LBGs appearing more frequently around bright foreground objects than in the total field, but the amplitude of the excess is much lower for the same flux limits of LBGs. However, for brighter flux limits, we see identical behaviour to that observed at higher redshift.}
\label{figure:proximity}
\end{figure*}

In this appendix, we present the manifestation of magnification bias in spatial correlations between bright foreground objects and bright LBGs, which illustrates the effect without relying on the Faber-Jackson relation.\\
\indent Source galaxies that have been magnified through gravitational lensing are necessarily located in close proximity to massive foreground objects. For the lensed fractions presented in Section \ref{section:results} we expect there to be an excess density of bright LBGs around bright foreground objects over the average field density. As the lensed fraction decreases with decreasing luminosity, the excess probability around bright deflectors should also decrease. Similarly, the clustering around the more massive, brighter deflectors should be stronger than around less massive, fainter deflectors.\\
\indent We compute the excess probability of finding an LBG brighter than $m=30, 28.5, 27.5$ \& $26.5$ at $\overline{z}=7.2$ within $5\farcs0$ of deflectors brighter than some $M_{B}$. We choose $5\farcs0$ as our limit as this is approximately the image-deflector separation beyond which strong lensing is unlikely. This is confirmed by the distribution of separations shown in Figure \ref{figure:lensprops}. We also present the spatial correlations for the lower redshift samples. At $\overline{z}=5.9$ we examine the same flux limits, at $\overline{z}=4.9$ we replace $m_{\textrm{lim}}=30$ with $m_{\textrm{lim}}=25.5$, and at $\overline{z}=3.8$ we replace $m_{\textrm{lim}}=28.5$ with $m_{\textrm{lim}}=24.5$. The results are plotted in Figure \ref{figure:proximity}.\\
\indent We find a large enhancement in the probability of finding bright LBGs nearby bright deflectors. As we consider fainter LBGs, the excess probability decreases monotonically in all of the samples. There exists a considerable excess of LBGs around foregrounds with $M_{B}<-23$ at $z\geq 4$, even for flux limits well beyond $M_{\star}$, however this signal is driven mainly by the brightest LBGs. The excess likelihood of locating an LBG around a foreground approaches unity by $M_{B}\sim-21$ for all flux limits in all samples.\\
\indent The clustering of bright LBGs nearby massive foreground galaxies is difficult to explain in the absence of magnification bias. One mechanism that could produce such a signal is the enhancement of LBGs around bright, red foregrounds. As discussed in Section \ref{subsection:contaminants}, we searched for this effect in LBG samples with spectroscopic follow-up from the literature, and found no evidence that bright, red foregrounds enhance LBG detection. Therefore, we conclude that the proximity effect shown in Figure \ref{figure:proximity} is consistent with being due to gravitational magnification of background LBGs by massive, bright foreground objects.\\

\label{lastpage}
\end{document}